\documentclass[12pt,a4paper]{article} 
\usepackage{graphicx} 
\usepackage{setspace}
\usepackage[paper=a4paper,left=30mm,right=30mm,top=30mm,bottom=30mm]{geometry}
\usepackage[footnotesize]{caption}
\begin{document}
%%%%%%%%%%%%%%%%%%%%%%%%%%%%%%%%%%%%%%%%%%%%%%%%%%%%%%%%%%%%%%%%%%
\renewcommand{\thefootnote}{\fnsymbol{footnote}}
\thispagestyle{empty}
\vspace*{1cm}
\begin{center}
 \section*{\bf A multi-agent-based approach to tax morale}
\vspace{0.5cm}

Georg Zaklan\footnote[4]{Department of Economics, University of Bamberg, Feldkirchenstrasse 21, D-96045 Bamberg, Germany.} \\
F. W. S. Lima\footnote[3]{Departamento de F\'{i}sica, 
Universidade Federal do Piau\'{i}, 64049-550 Teresina -- PI, Brazil.}\\
Frank Westerhoff$^{\S}$\\[0.3cm]
\vspace{0.5cm}
e-mail:   georg.zaklan@uni-bamberg.de\\[1cm]   %wel@ufpi.br; 
\vspace{0.5cm}
\today 
\end{center}

\vspace{0.55cm} 
\begin{abstract}

\noindent We embed the behaviour of tax evasion into the standard two-dimensional 
Ising model. In the presence of an external magnetic field, the Ising model is 
able to generate the empirically observed effect of tax morale, i.e. the 
phenomenon that in some countries tax evasion is either rather high or low. 
The external magnetic field captures the agents' trust in governmental institutions. 
We also find that tax authorities may curb tax evasion via appropriate enforcement 
mechanisms. Our results are 
robust for the Barab\'asi-Albert and Voronoi-Delaunay networks.\\[0.3cm]
\noindent [Keywords: Opinion dynamics, 
Sociophysics, Ising model, Tax morale.]\\[0.0cm]
%\textbf{JEL classification}
\end{abstract}
\newpage
\renewcommand{\thefootnote}{\arabic{footnote}}
\section{Introduction}
Tax evasion can vary widely across nations, reaching extremely high values in some
developing countries. Wintrobe and G\"erxhani (2004) explain the observed higher
levels of tax evasion in generally less developed countries with a lesser amount of
trust that people accord to governmental institutions. Many empirical studies 
confirm that tax payers are more willing to comply, the more they feel that the 
government represents their preferences, i.e. the more  content they are with how 
their government uses tax revenue (see, e.g.: Schnellenbach, 2002; Torgler, 2004; 
Hyun, 2005; Cummings et al., 2007). We consider the concept of tax morale important 
because it seems for the most part to explain why compliance differs across 
countries even if the level of enforcement is roughly the same. 
Our aim is to add to the experimental and empirical studies conducted on tax morale 
by applying a multi-agent-based approach. In economics, the problem of tax evasion 
from a multi-agent-based perspective has received little attention to date (see 
Bloomquist (2006) for a recent overview).

In two previous studies (Zaklan et al., 2008; Westerhoff et al., 2008) we modify 
the two-dimensional Ising model to analyse how enforcement affects tax evasion 
dynamics. The key idea in these papers is that agents may either behave honest 
(tax payers) or dishonest (tax evaders), a decision which is subject to group influence. 
We further augment the Ising model by adding an enforcement mechanism, which consists 
of two elements: a probability of an audit ($p$) and a period of honesty ($k$) in which 
detected evaders remain honest. We find evidence that enforcement has a direct and 
indirect effect on aggregate tax evasion. First, enforcement obviously has a direct 
effect on aggregate tax evasion since any tax evader who is audited becomes honest for 
a certain period of time. Second, there is an important indirect effect of enforcement 
on aggregate tax evasion since agents who live in an environment with more honest agents 
are also likely to be honest, due to group influence. We also find that even low levels 
of enforcement may help to prevent substantial fluctuations in tax evasion.

The aim of the current work is to develop our previous studies by showing that our tax 
evasion model can also replicate the effect of tax morale, if an external magnetic field 
is incorporated. By modifying the probabilities of evading or not evading tax duties, the 
external magnetic field captures agents' trust in governmental institutions.
By implementing a positive and a negative external magnetic field and contrasting the 
resulting equilibrium levels of tax evasion with the baseline setting, where the external 
magnetic field is absent, 
we show that our model can explain both high and low levels of tax morale. To demonstrate 
that this remains valid regardless of the level of enforcement, we depict the equilibrium 
values of tax evasion corresponding to each considered external magnetic field for 
different levels of enforcement.

The remainder of our manuscript is organised as follows: in section 2 we briefly present
our tax evasion model, which is based on the standard two-dimensional Ising model on a 
square lattice. In section 3, we discuss how the external magnetic field influences tax 
evasion. By varying the compliance level systematically and observing the resulting 
equilibrium levels of tax evasion for different external magnetic fields, we illustrate 
that the external magnetic field may be an appropriate tool which enables different 
degrees of tax morale to be modeled, for any level of compliance.   
In section 4, we also embed our model into the Barab\'asi-Albert network and the 
Voronoi-Delaunay random lattice. We find that the effect of the external magnetic 
field also remains valid in these networks.
The last section presents a number of conclusions.

\section{The model}
We use the so-called single spin-flip heat-bath algorithm to simulate the Ising model 
on a 1000 $\times$ 1000 square lattice with periodic boundary conditions. In this 
algorithm, irrespective of the current state of the chosen spin, its new state is chosen 
with probabilities 
$P_{\pm}=\mbox{e}^{-\beta E_\pm}/(\mbox{e}^{- \beta E_{+}} + \mbox{e}^{- \beta E_{-}})$, 
where $P_\pm$ are the probabilities of choosing the up ($+$) and down ($-$) states, 
respectively (Chowdhury and Stauffer, 2002), and $\beta$ is defined as $1/(k_B \cdot T)$. 
The total energy ($E$) is given by the Hamiltonian $H=-\sum_{<i,j>}J_{ij} S_i S_j-B\sum_{i}S_i$ 
(we use $J_{ij}=J=1$). As an external magnetic field we use $h=2B/k_B T$. A negative external 
magnetic field acts to augment the probability of becoming dishonest, whereas a positive field 
lowers this probability. In our previous models we set $B=0$, and hence neglect the influence 
of the external magnetic field. We interpret a negative external magnetic field ($h<0$) as 
agents' low confidence in govenmental institutions, meaning that tax evasion  is consequently 
high.
On the other hand, a positive external magnetic field ($h>0$) implies  above-average confidence 
in the tax authorities: since individuals feel well represented by the government, they feel 
more comfortable about paying their tax duty, and hence evade paying tax less frequently.

We interpret this setup in the following way: in every time period each lattice site is occupied 
by a tax payer who can either be honest ($S_i = +1$) or a tax evader ($S_i =-1$). We assume that  
everybody is honest, initially. In consecutive time periods agents have the opportunity to change 
their agent-type, enabling cheating agents to become honest and honest citizens to become tax 
evaders. Each agent's social network is made up of four nearest neighbours, and they may either 
prefer tax evasion, reject it or be indifferent.

Tax evaders have the greatest influence to turn honest citizens into tax
evaders if they constitute a majority in the given neighbourhood. 
If the majority evades paying tax, one is likely to also evade. On the other
hand, if most people in the vicinity are honest, the respective individual is 
likely to become a decent citizen if she was a tax evader before. The strength 
of neighbourhood influence can be controlled by adjusting the temperature, $T$.  

For very low temperatures, individuals mainly base their decision regarding tax 
evasion on what the neighbourhood does. A rising temperature has the opposite 
effect. Individuals then decide more autonomously. 
We only use temperatures below $T_c$ ($\approx 2.269$)  for our simulations, 
since we are interested in modeling the effect of group influence. 

As an enforcement measure, we introduce a probability of an efficient audit ($p$). 
If tax evasion is detected, the individual remains honest for a certain number of 
periods. We denote the period of time for which detected tax evaders remain honest 
by the variable $k$. We assume that $k$ is a random number between either 0 and 10 
or 0 and 50 periods. Both of these intervals, which we will use, express that agents 
are ashamed and feel guilty about their behaviour if cheating is detected, but that 
the shame is of temporary nature, which is typical for the normal type of shame 
(e.g. Potter-Efron and Efron (1993)). 
Obviously, the second interval allows detected agents to be more ashamed on average, 
having evaded taxes: agents have a longer memory of their wrongdoing.

One time unit is one sweep through the entire lattice. Audits are stochastically 
independent from other agents and from the history any agent has.

\section{Dynamics of the model}

In Figure 1, the probability of an audit is set to 1\% and the temperature to a 
level 25\% below $T_c$. The upper left panel in Figure 1 ($k\in [0,10]$ and $h=-0.25$) 
illustrates that tax evasion may spread throughout the population within only 1000 time 
steps, given that the interval for the parameter $k$ is chosen to be sufficiently small. 
The level of tax evasion comes to rest at about 90 percent. The three panels below, also 
in the first column of Figure 1, visualise the distribution of agent types after 50, 500 
and 1000 time steps, respectively. It can be seen that both honest people (white) and 
evaders (black) increasingly group together in clusters as time passes. After 50 time 
steps, when the aggregate tax evasion is at about 4 percent, the clusters are still 
relatively small. After 500 time steps (about $35\%$ of the population then evades taxes), 
the clusters grow to such an extent, that individuals with different opinions regarding tax 
evasion clearly separate. After 1000 time steps, tax evasion eventually prevails, as the 
diagram at the bottom left clearly illustrates.

\vspace{0.3cm}
\begin{center}
---$\quad$Figure $1$ goes about here$\quad$---
\end{center}
\vspace{0.3cm}

\noindent The column on the right of Figure 1 shows that higher values of $k$ ($k\in[0,50]$) 
prevent this spread of tax evasion, keeping it at about 4\% over time. The other three diagrams 
in the right-hand column also illustrate the distribution of agent types after 50, 500 and 1000 
time steps, respectively, and show that honest individuals remain dominant over time.

The left column of Figure 2 also implies a maximum duration of honesty of 10 periods, whereas 
the right column shows a maximum duration of honesty of 50 periods, after having been subject 
to an audit. 
We analyse for different strengths of group influence (by systematically lowering the 
temperature further below $T_c$, we accord more importance to group influence) how the 
equilibrium levels of tax evasion evolve under different influences of the external 
magnetic field for different enforcement levels: we implement these by gradually augmenting 
the probability of an audit from 0 to 5\% (in steps of 0.1 percent).

\vspace{0.3cm}
\begin{center}
---$\quad$Figure $2$ goes about here$\quad$---
\end{center}
\vspace{0.3cm}

\noindent Each equilibrium value in tax evasion is calculated by first allowing a transient 
phase of 5000 time steps and by then forming the simple average of the tax evasion levels over 
the next 1000 time steps.  In Figure 2 the lines are marked identically to denote the same 
influence of the external magnetic field: diamonds imply $h=-0.5$, squares denote the absence 
of the external field (i.e. $h=0$)
and pluses represent the case where the magnetic field is positive ($h=0.5$).  

Figure 2 illustrates that a positive external field ($h=0.5$) works to lower tax evasion, compared 
to the baseline case, where an external field is absent. On the contrary, a negative external 
field works to augment the problem of tax evasion: the curve where $h=-0.5$ (diamonds) is above 
and the curve where $h=0.5$ (pluses) is below the line where $h=0$ (squares), for all considered 
probabilities of an audit. Recall that a positive value of $h$ stands for the case in which agents 
have great trust in governmental institutions and that a negative value of $h$ stands for the case 
in which agents distrust governmental institutions. In this sense our model is able to replicate the 
phenomenon of tax morale.

Note that, if the external magnetic field is negative, an increase in the audit rate has the effect 
of reducing non-compliance substantially. Given that individuals are sufficiently affected in the 
case of detection (right-hand column of Figure 2) and group influence is sufficiently strong (panel 
on the bottom right), non-compliance may be reduced from about $100\%$ to below $10\%$ by increasing 
the audit rate to 5\%.

\section{Modifications}
We now consider our model on other network structures to obtain further support for our results. 
In particular, we use the Voronoi-Delaunay random lattice and the Barab\'asi-Albert network model. 
The Voronoi-Delaunay lattice (i.e. tessellation of the plane for a given set of points) is 
constructed as follows (Lima et al., 2000). First, a given number of points are randomly 
distributed in a plane of a given size. For each point, the polygonal cell, consisting of 
the region of space nearer to that point than to any other point, must be determined. Whenever 
two such cells share an edge, they are considered to be neighbours.
Using the Voronoi tessellation, the dual lattice can be obtained as follows: if two cells are 
neighbours, a link is placed between the two points located in the cells. The triangulation of 
space is obtained from the links. The network constructed in this way, which we use for simulation, 
is called the Voronoi-Delaunay lattice. 

The Barab\'asi-Albert network (Barab\'asi and Albert, 1999) is established such that the probability 
of a new site being connected to one of the already existing sites is proportional to the number of 
connections the existing site has already accumulated over time: the chance for an agent to obtain 
a new connection is greater if he is well connected already.
 
In these variations of our simple square lattice model, we also choose 1,000,000 agents and calculate 
the equilibrium levels of tax evasion. The results are displayed in Figures 3 and 4.
As can be seen in both networks, the Voronoi-Delaunay lattice (Figure 3) and the Barab\'asi-Albert 
network (Figure 4), higher values of $k$ (i.e. $k\in[0,50]$ versus $k\in[0,10]$) also lead to lower 
equilibrium values in tax evasion in any considered parameter setting, as in the square lattice 
network: identically marked curves in panels on the right are lower than the corresponding curves, 
which are in the same row, but on the left.

\vspace{0.3cm}
\begin{center}
---$\quad$Figures $3$ and $4$ go about here$\quad$---
\end{center}
\vspace{0.2cm}

\noindent In the Voronoi-Delaunay network (Figure 3), the curves, where the external magnetic 
field is absent ($h=0$) and where it is positive ($h=0.5$) resemble much the corresponding 
lines in the square lattice network. Yet enforcement seems to work better. Especially when 
looking at the case where the external magnetic field is negative ($h=-0.5$),
one can clearly see that non-compliance may be reduced from very high to quite low levels. 
For example, if group influence is strong ($T=0.8 \cdot T_c$) and the agent's memory 
typically high ($k\in[0,50]$), the influence of the negative external magnetic field, 
which favors non-compliance, may quickly be reduced by augmenting the probability of an 
audit to such low levels as 0.5\%.

Tax evasion is quite different under the Barab\'asi-Albert network (Figure 4). If a negative 
external field is applied ($h=-0.5$), tax evasion is lower for small audit rates than in the 
square lattice or in the Voronoi-Delaunay network. While tax evasion increases to nearly 100\% 
in the other two networks if no audits are conducted, in the Barab\'asi-Albert network, on the 
other hand, non-compliance only rises to about 70\% (for $p=0$).
While tax evasion is lower for small audit rates in the Barab\'asi-Albert network, if a 
negative external field is applied, in the case of no external field ($h=0$) or a positive 
one ($h=0.5$), equilibrium tax evasion is much higher than in the other two networks. Also 
audit rates have a different impact on tax evasion, compared to the other networks. While in 
the square lattice and the Voronoi-Delaunay lattice tax evasion under a zero external and a 
positive external magnetic field are only marginally reduced, in the Barab\'asi-Albert network 
tax evasion can be reduced substantially, by increasing the audit rate to 5\% (here we mean the 
amount of more compliance that can be established by implementing higher levels of enforcement). 
If a negative external field is applied, it can be seen, as in the other two networks, that tax 
evasion decreases with an increasing audit rate and that for any level of enforcement (i.e. audit 
rate) non-compliance is higher than for a zero or a positive external magnetic field.

For sufficiently small enforcement levels, we find for the applied negative external magnetic 
field ($h=-0.5$) that the equilibrium values 
in tax evasion are higher, the further the temperature is decreased.\footnote{However, if the 
temperature is below a certain treshold (i.e. about $T=0.55\cdot T_c$ for the square lattice 
with $k\in[0,10]$), the negative external magnetic field we apply is too small, to trigger 
non-compliance to spread. For sufficiently low temperatures the state of compliance hence 
prevails for the considered negative external magnetic field, even for low audit rates.} 
This is most obvious for the square lattice with $k\in[0,10]$: 
low enforcement levels (i.e small values of $p$) are less efficient to reduce tax evasion, 
the more individuals focus on what their neighbours do (i.e. for lower temperatures). In the 
presence of the negative external magnetic field, the number of honest agents is small for 
low enforcement levels.  
At lower temperatures honest agents (some agents become honest due to enforcement) have less 
influence on cheating agents to become honest, so that, compared to higher temperatures, tax 
evasion is higher for low enforcement levels. 
If enforcement levels increase, the tax evasion of more people is detected and these agents 
then usually become compliant for a while. If the number of agents is sufficiently large 
(due to a high level of enforcement), their influence on the non-compliant individuals 
grows. The lower the temperature is, the greater is their influence: non-compliance hence 
is smaller at lower temperatures, if enforcement is sufficiently high. This can be observed 
also in the Voronoi-Delaunay network for $k\in[0,10]$ and in the 
Barab\'asi-Albert network for lower values of $k_{max}$, e.g. $k_{max}=5$.

Finally we briefly discuss why the non-compliance curve, which captures the influence of 
the negative external magnetic field ($h=-0.5$), declines more rapidly in the Barab\'asi-Albert 
and in the Voronoi-Delaunay network, compared to the square lattice, when the audit 
rate is increased.
We conjecture that this can be explained by the greater mean size of the neighbourhood 
in these two networks (on average, the neighbourhood comprises 6 agents in the 
Barab\'asi-Albert network and 8 agents in the Voronoi-Delaunay lattice), compared to the square 
lattice (each agent has only 4 nearest neighbours).  Setting the audit rate to a positive value 
forces some agents in the neighbourhood to become honest for $k$ periods. The greater the size 
of the neighbourhood is, the more honest agents will be included in the considered agent's 
social network. These honest agents then exert influence on the respective individual to become 
honest, too. Hence, in the Voronoi-Delaunay lattice, which on average includes most agents in a 
neighbourhood, increasing the audit rate slightly has the effect of reducing non-compliance the 
most. In the Barab\'asi-Albert network, whose neighbourhood is intermediate in size, the same 
increase of the audit rate consequently works less effectively to establish more compliance, 
compared to the Voronoi-Delaunay lattice, although it is still better than in the square lattice.

\section{Conclusion}
By incorporating the possibility of tax evasion into the Ising model we are able to replicate, 
within the context of a multi-agent-based model, the empirical fact regarding the existance 
of tax morale by isolating the effect of the external magnetic field on tax evasion. While 
in this study our focus is on
the influence of the external magnetic field on tax evasion, other 
extensions to our simple model appear interesting and economically important. Allowing agents 
to differ in their period of honesty after detection seems to be just the first step in 
creating a more realistic tax evasion model. Economic variables, which are important for 
the decision regarding tax evasion, such as personal income, may need to be incorporated 
into the model.
Also, an intuitive extension to our hitherto very simple model may be to allow for a third 
type of agent, who is neither entirely compliant nor entirely non-compliant, but somewhere 
inbetween.

\vspace{1cm}
\noindent {\bf Acknowledgements:} The authors are pleased to thank D. Stauffer for his 
helpful suggestions on how to implement an efficient heat-bath algorithm with an external 
magnetic field. Further, F. W. S. Lima acknowledges the support of the SGI Altix 1350 system 
at the computational park CENAPAD.UNICAMP-USP, SP-BRAZIL.

\newpage
%\section*{\hspace*{-2ex}Appendix A: Fortran source code for a single
%  time series}
%\setstretch {0.969}
%\small

%\lstset{language=Fortran}
%\lstinputlisting[numbers=left,breaklines]{quellcode.txt}

%=============================LITERATURE===============================
\newpage
\setstretch {1.5}
 
\newpage
\pagestyle{empty}

%%%%%%%%%%%%%%%%%%%%%%%%%%Abbildung 1%%%%%%%%%%%%%%%%%%%%%%%%%%
\newpage
\begin{center} 
\begin{tabular}{c}
%1
\includegraphics[width=6.5cm]{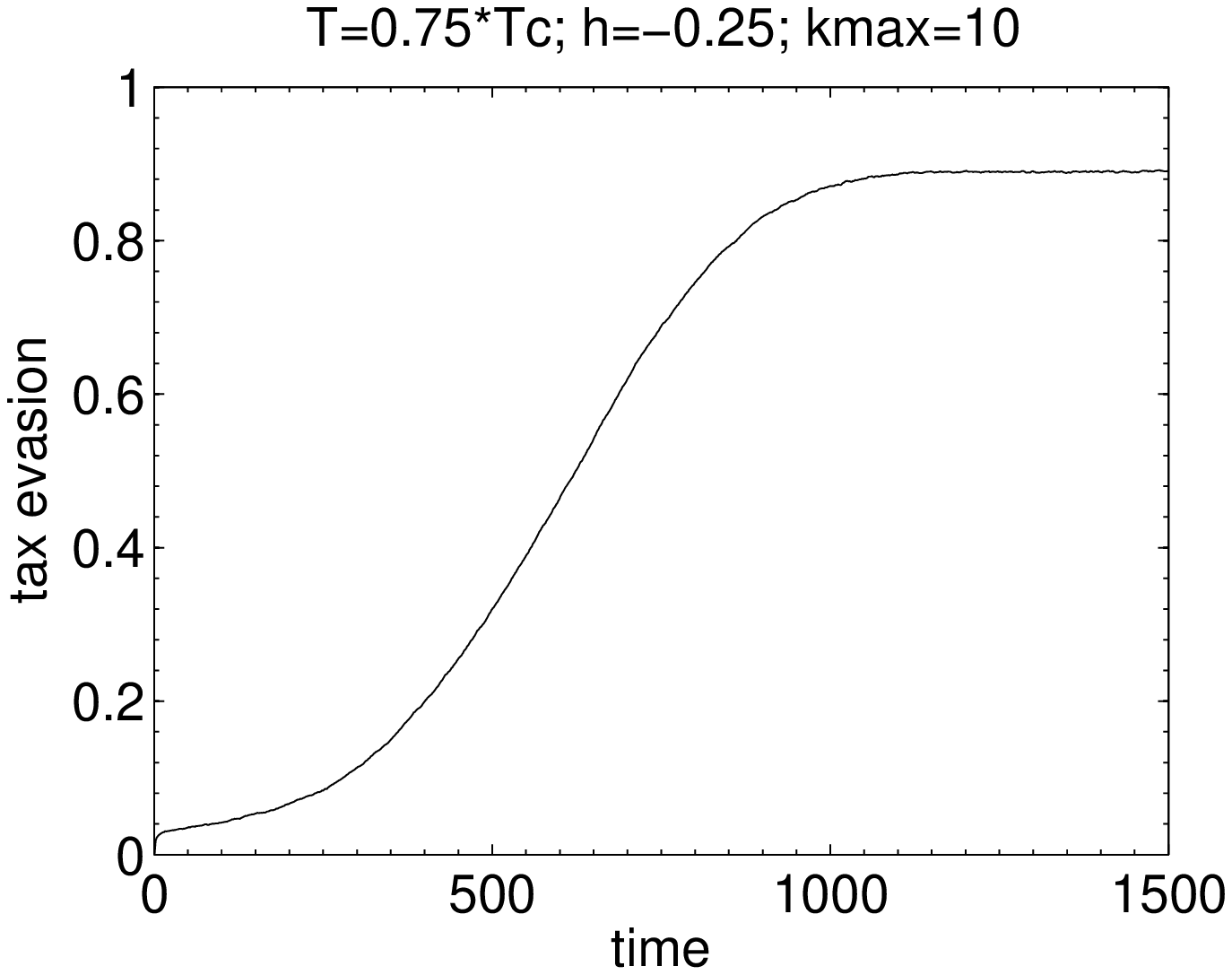}
\hspace{0.25cm}
\includegraphics[width=6.5cm]{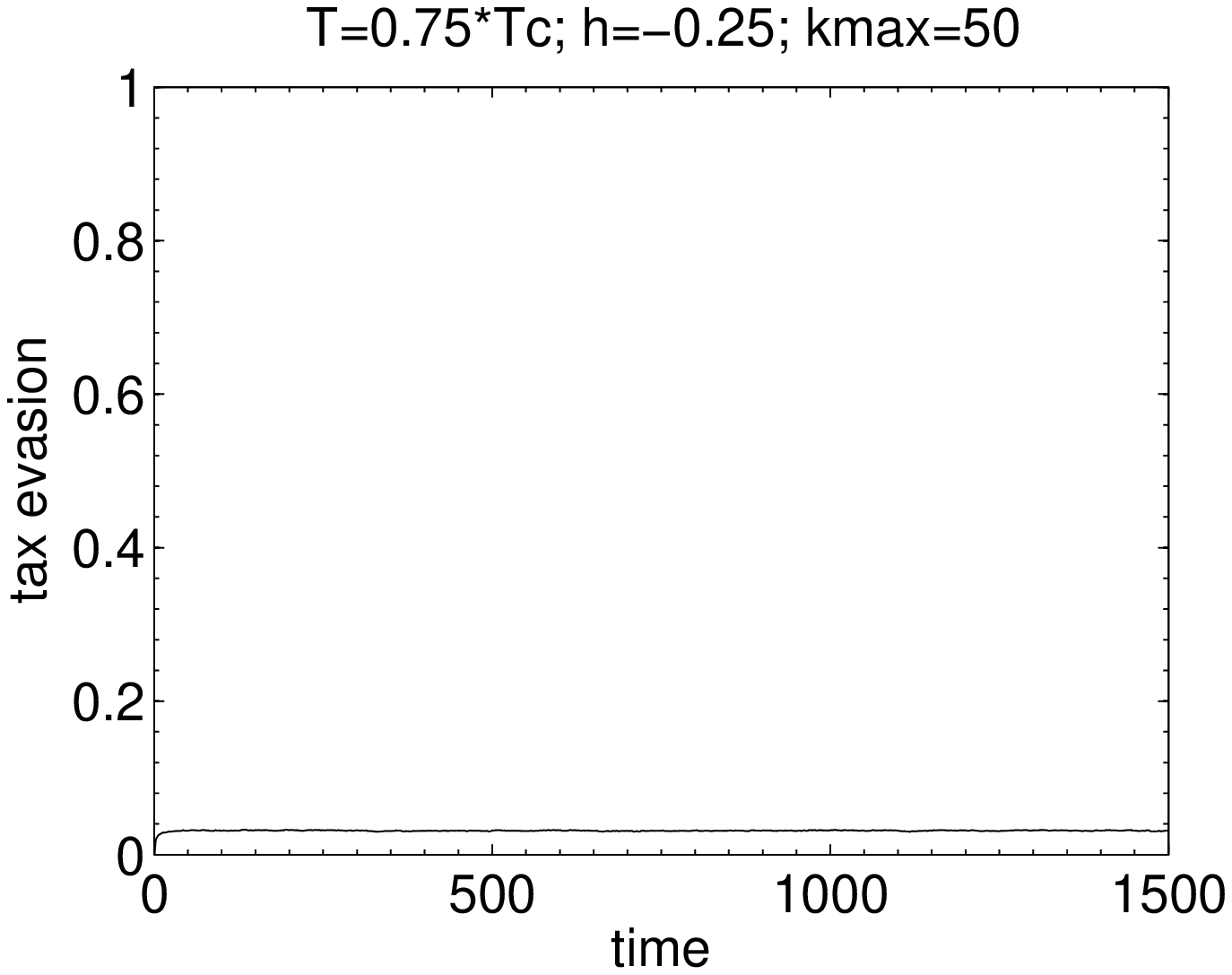}\\[0.25cm] 
%2 
\hspace{0.3cm}\framebox{\includegraphics[width=4.5cm]{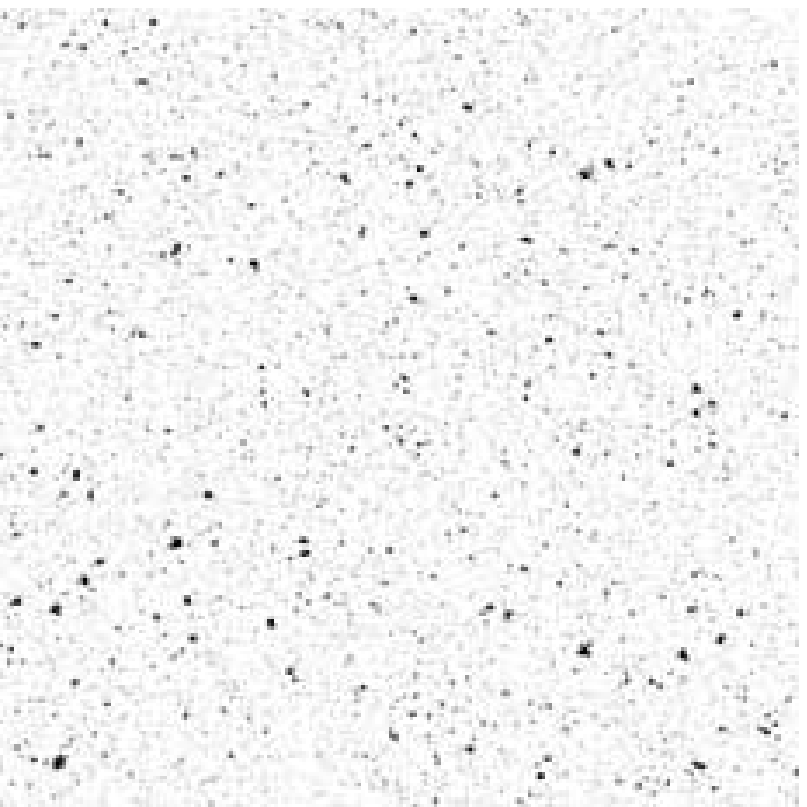}}
\hspace{2.0cm}
\framebox{\includegraphics[width=4.5cm]{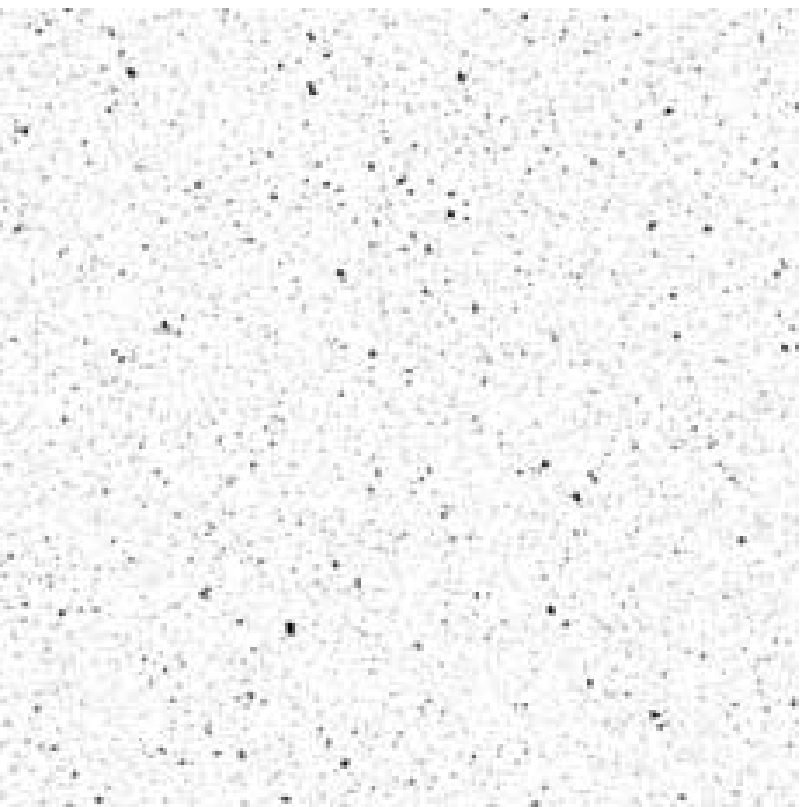}}\\[0.25cm]    
%3
\hspace{0.3cm}\framebox{\includegraphics[width=4.5cm]{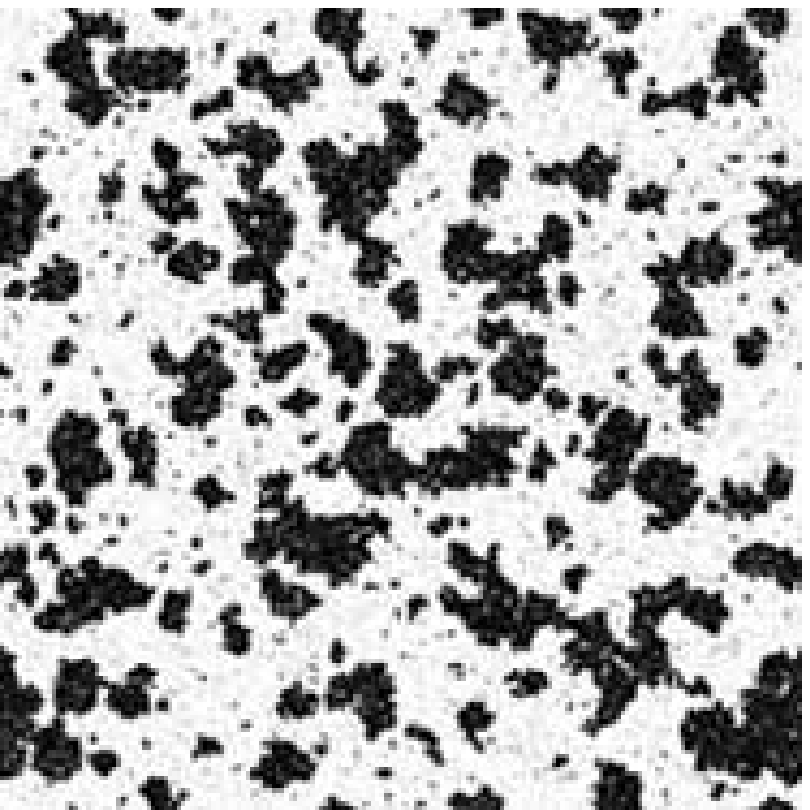}}
\hspace{2.0cm}
\framebox{\includegraphics[width=4.5cm]{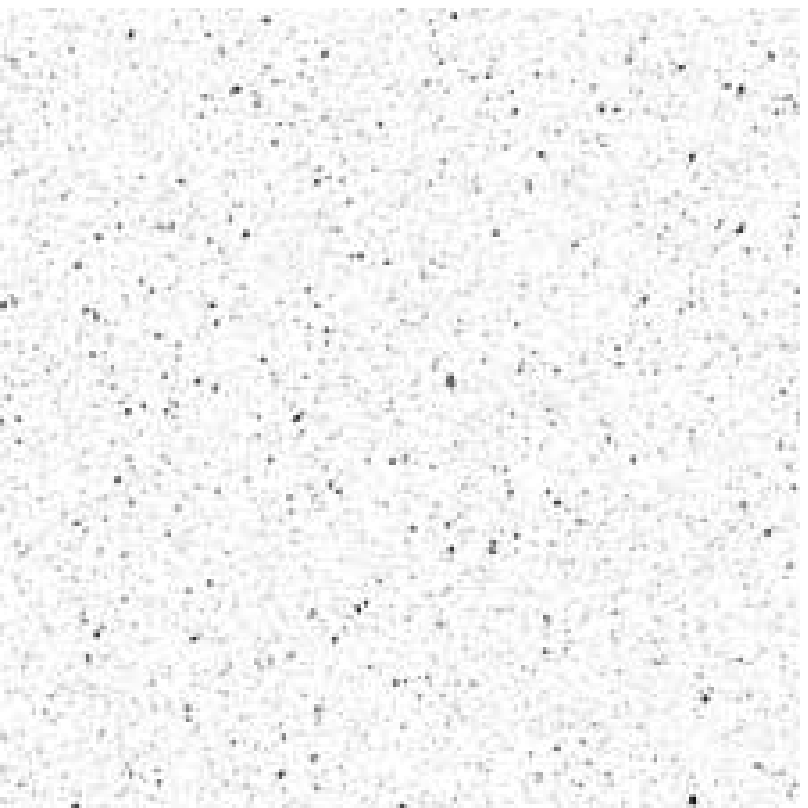}}\\[0.25cm]    
%4
\hspace{0.3cm}\framebox{\includegraphics[width=4.5cm]{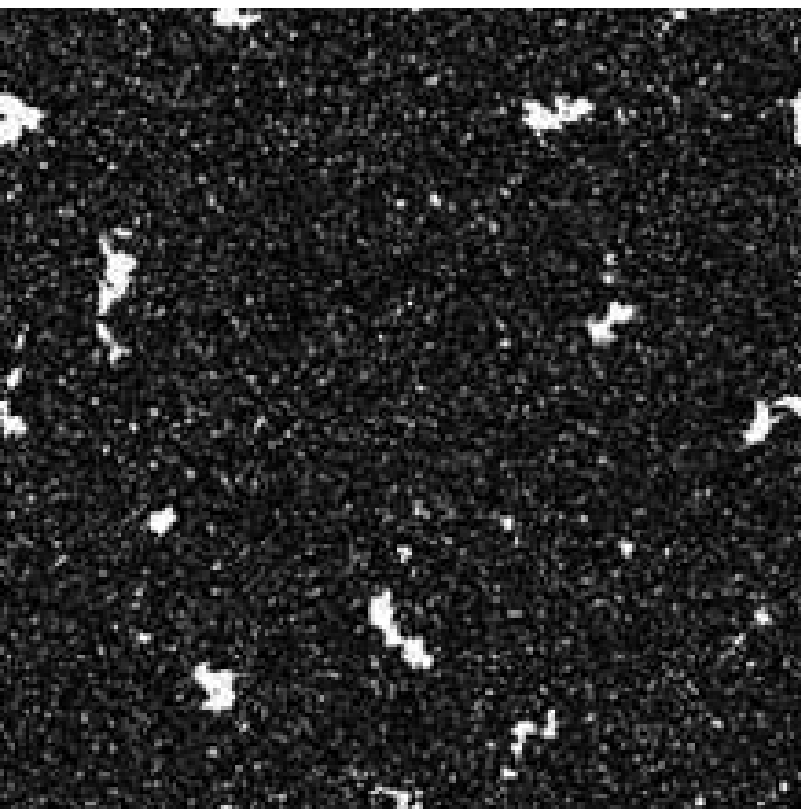}}
\hspace{2.0cm}
\framebox{\includegraphics[width=4.5cm]{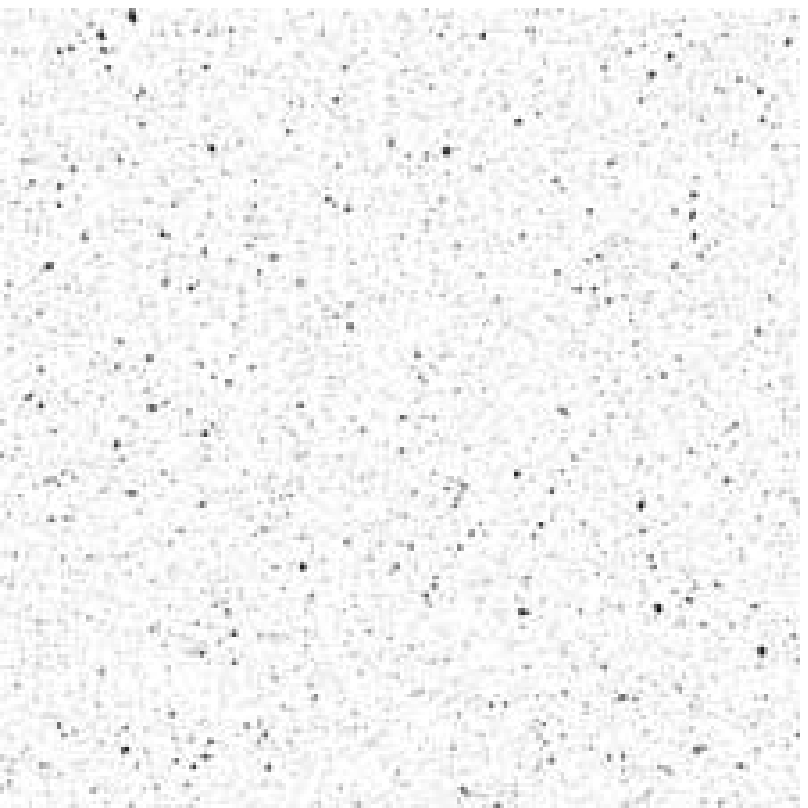}}
%ende
\end{tabular}
\end{center}

\enlargethispage{\baselineskip}
\enlargethispage{\baselineskip}
\begin{minipage}[t]{\textwidth}
\vspace{0.25cm}
Figure 1:  Illustration of how a negative external magnetic field (here we use $h=-0.25$) 
may cause most of society to become non-compliant. (A more detailed decription is 
contained in section 3.)
\end{minipage}

%%%%%%%%%%%%%%%%%%%%%%%%%%Abbildung 2%%%%%%%%%%%%%%%%%%%%%%%%%%5
\newpage
%\vspace*{-2.5cm}
\begin{center} 
\begin{tabular}{c}
\includegraphics[width=6.5cm]{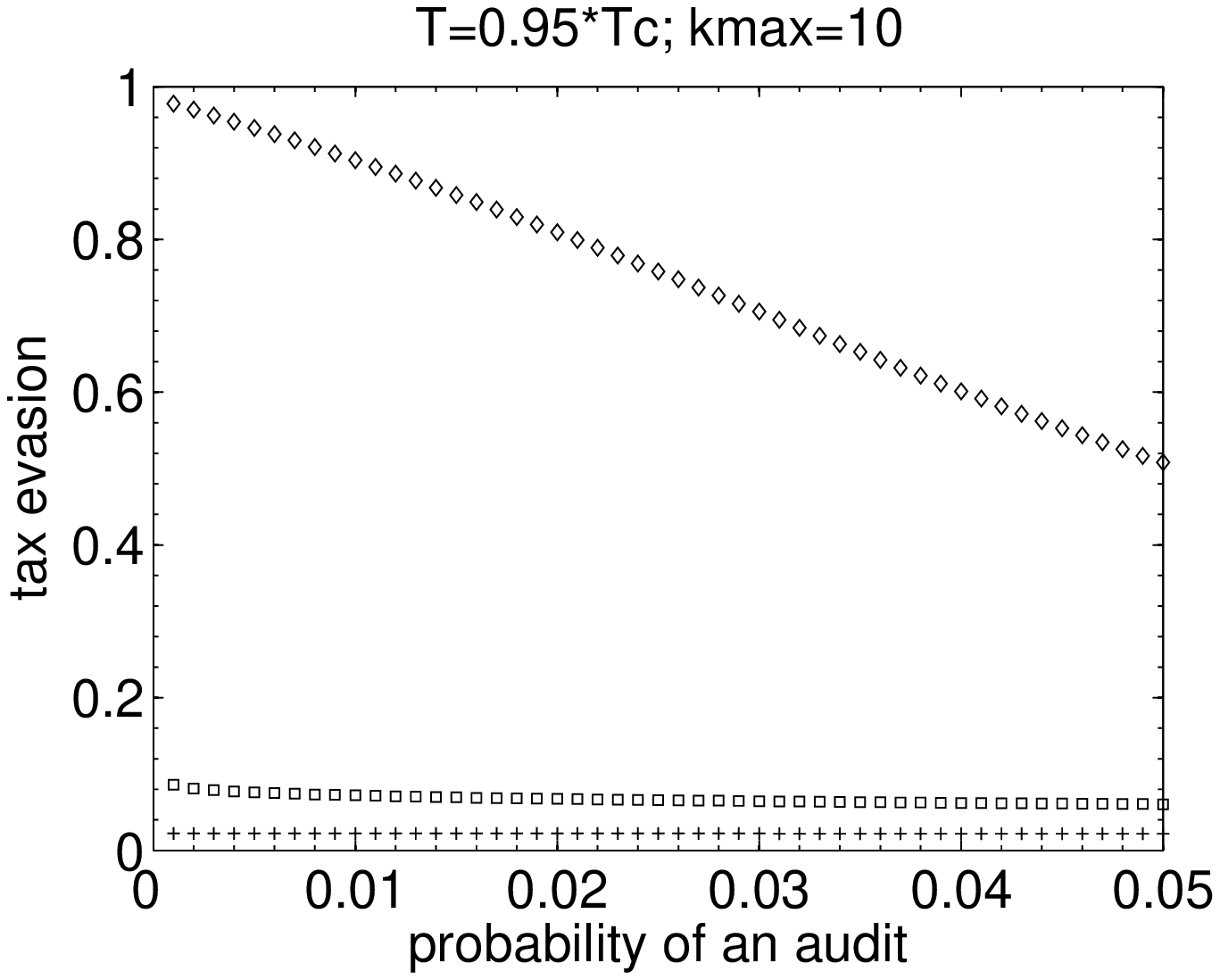}
\hspace{0.25cm}
\includegraphics[width=6.5cm]{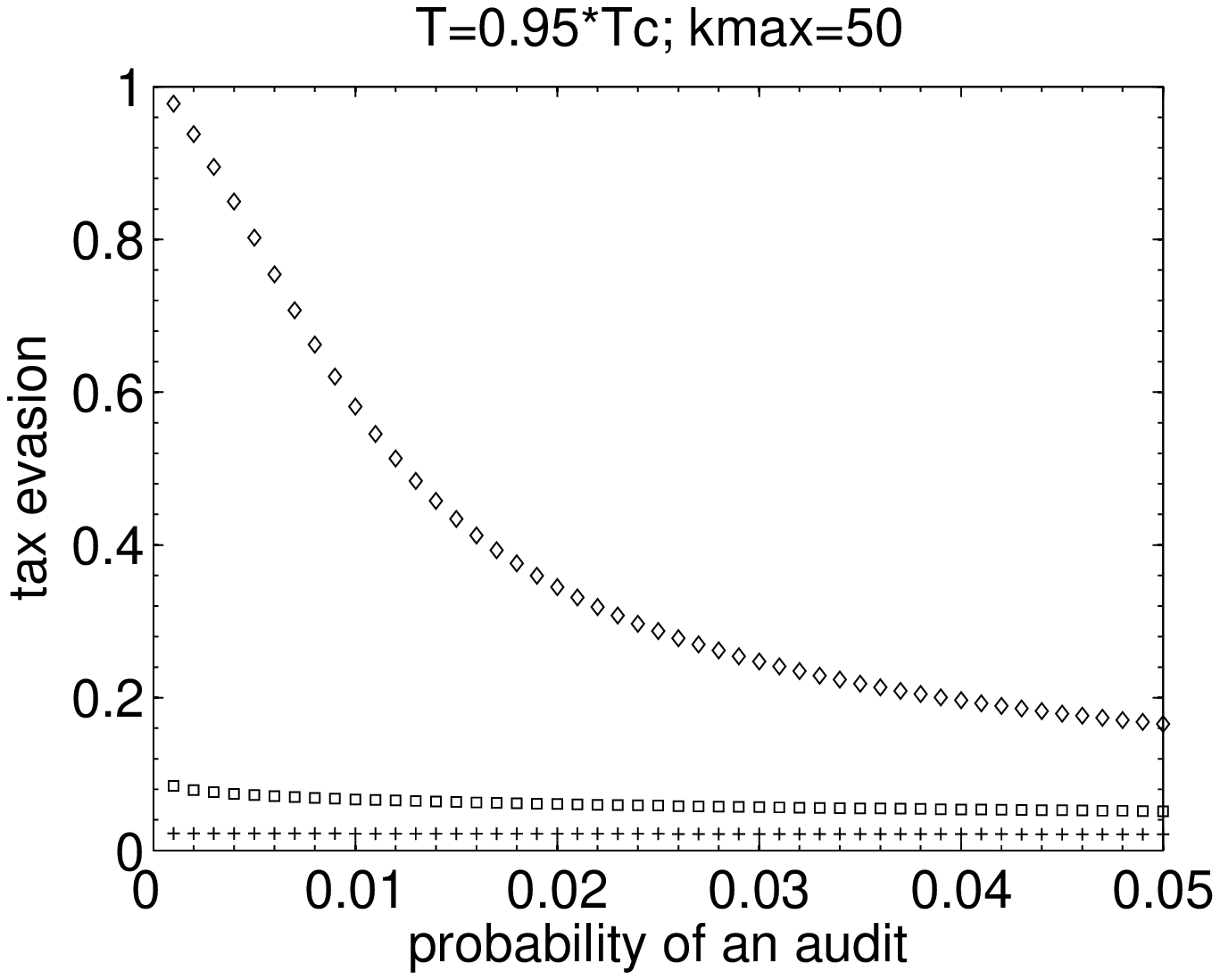}\\[0.25cm]  
\includegraphics[width=6.5cm]{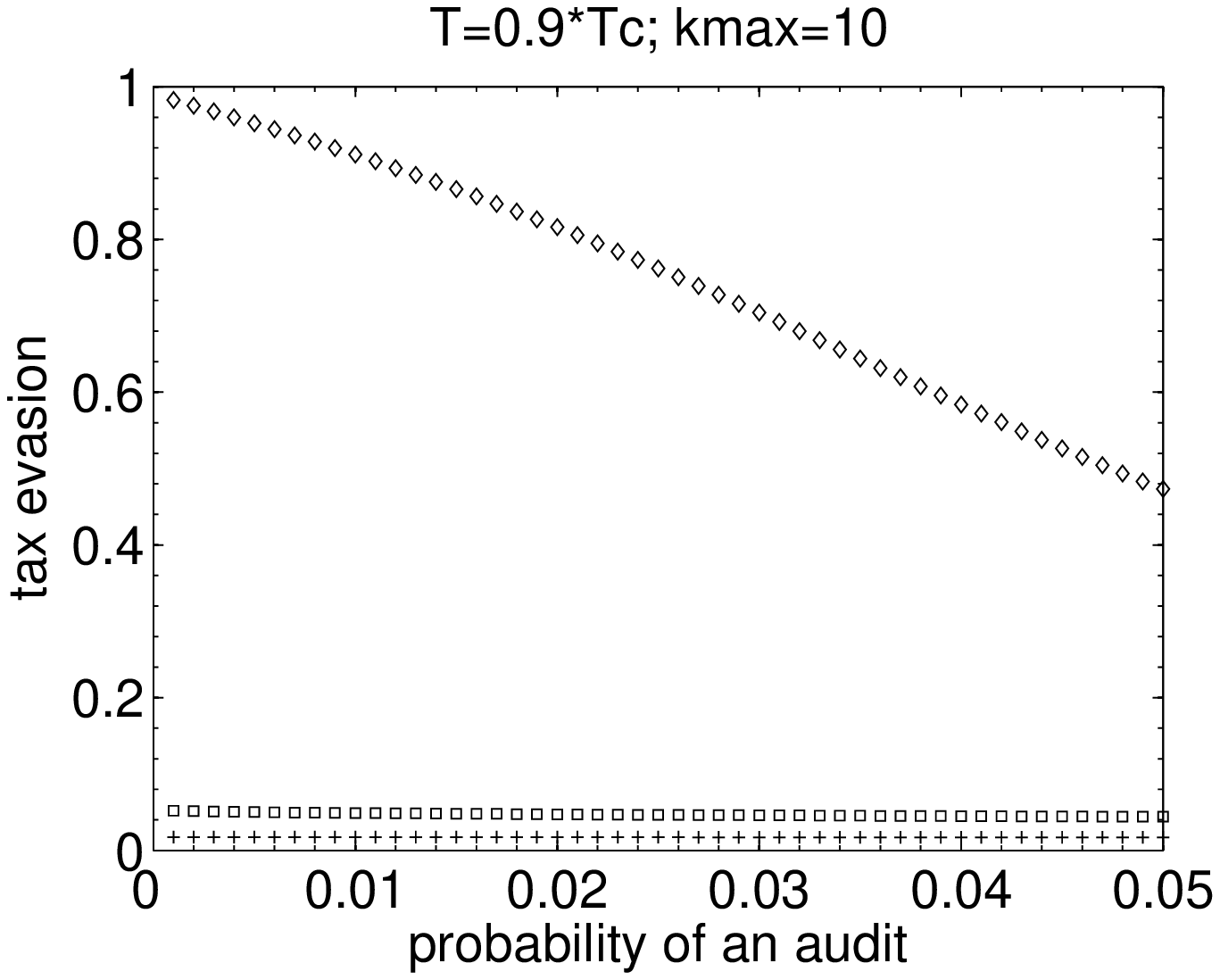}
\hspace{0.25cm}
\includegraphics[width=6.5cm]{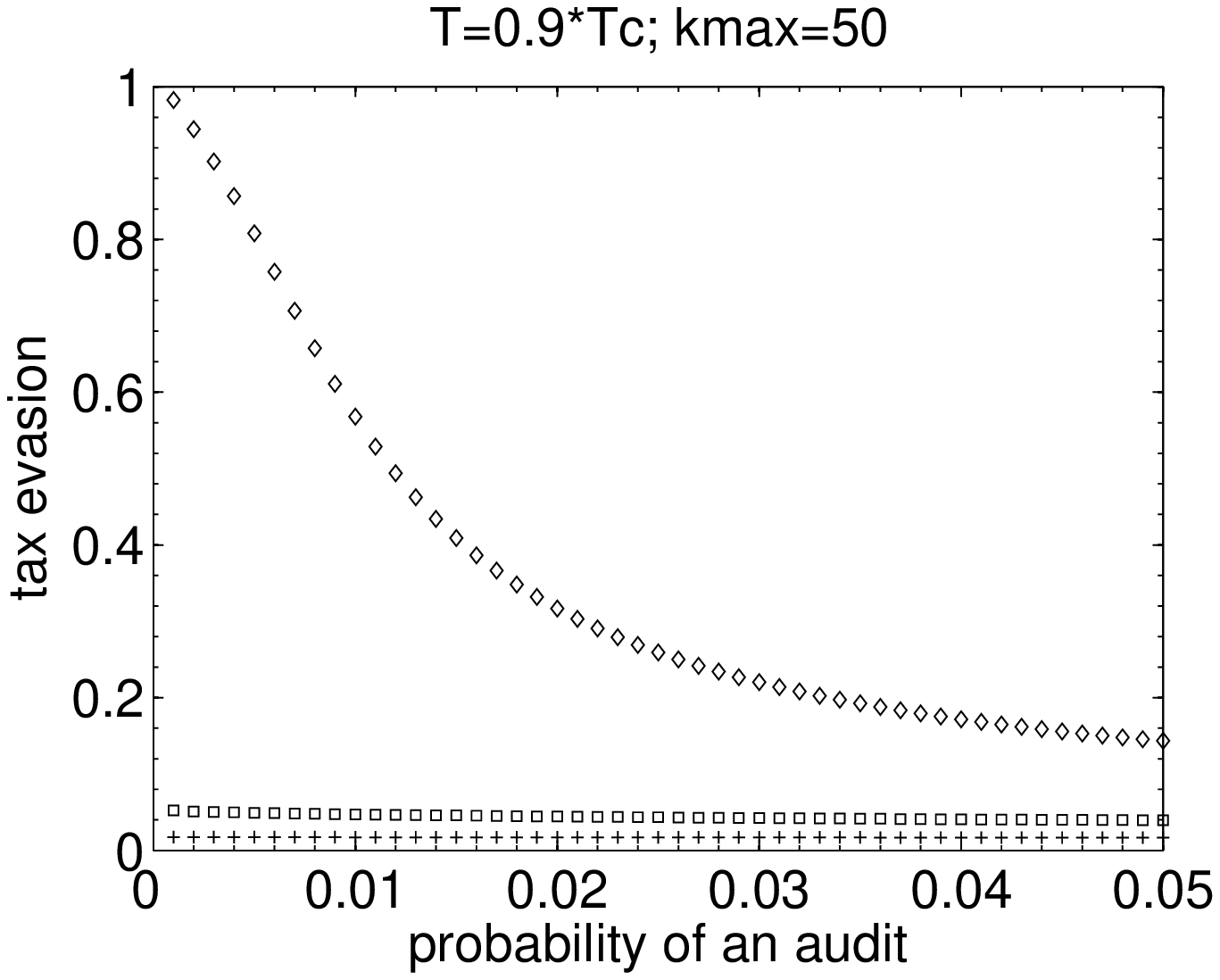}\\[0.25cm]    
\includegraphics[width=6.5cm]{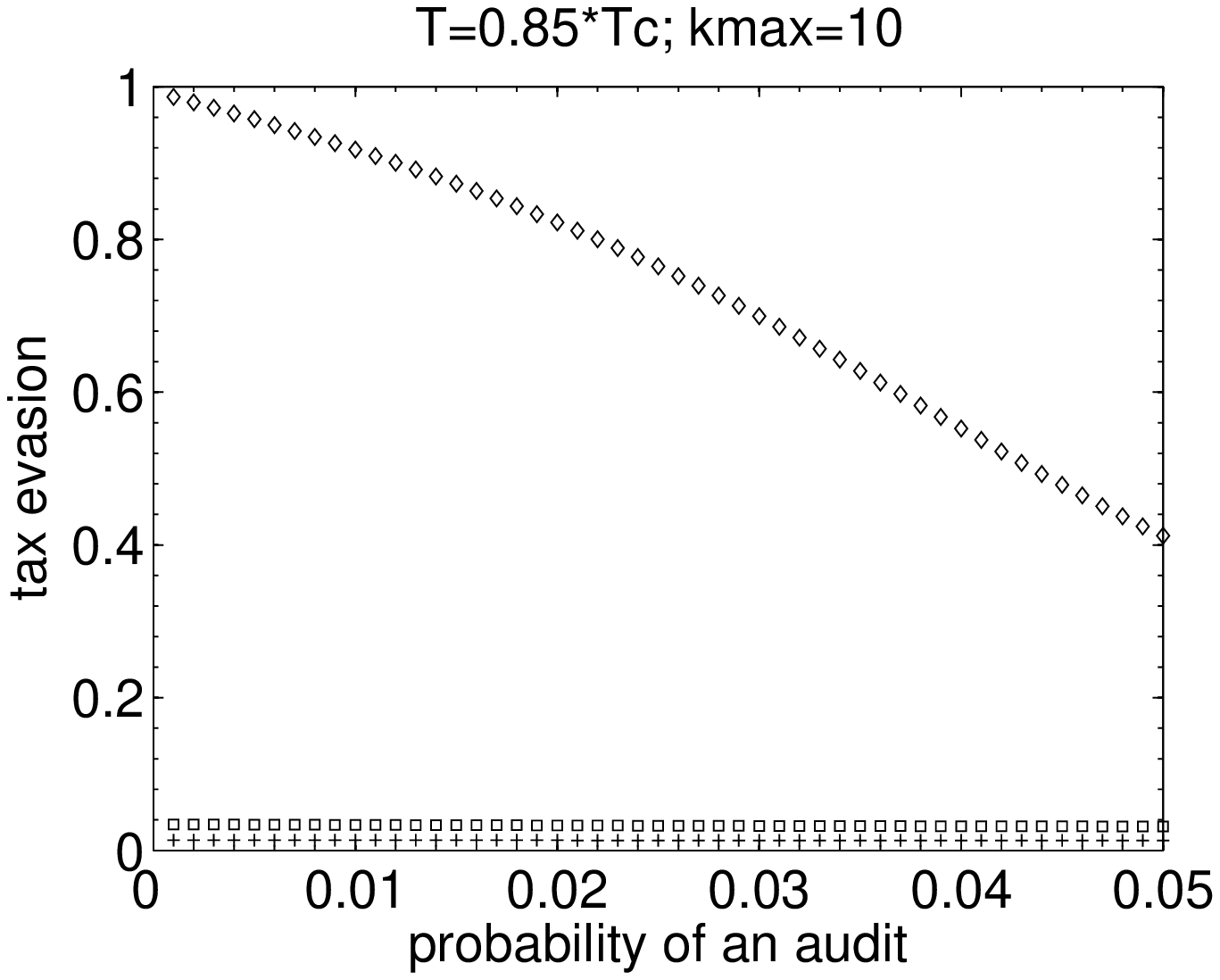}
\hspace{0.25cm}
\includegraphics[width=6.5cm]{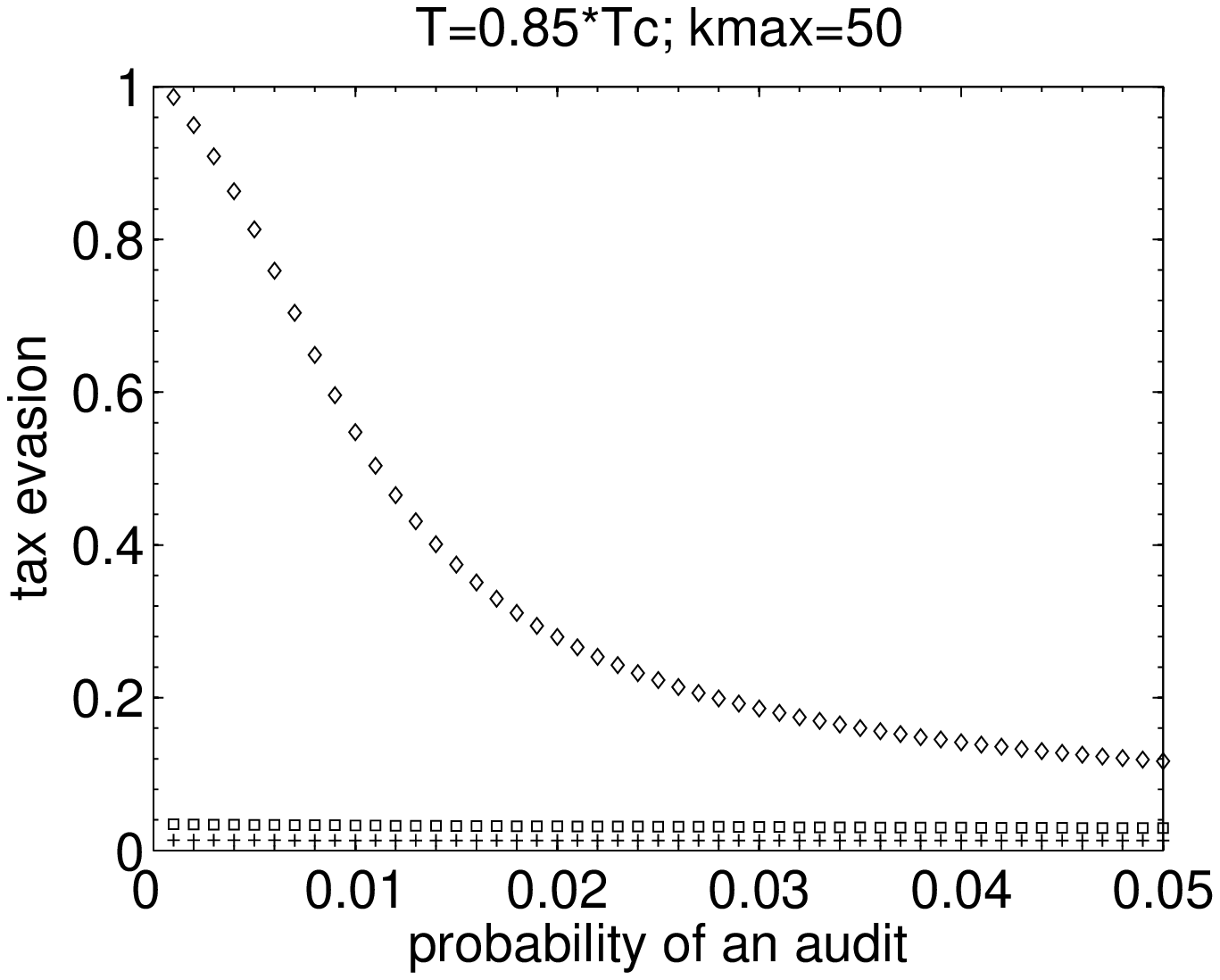}\\[0.25cm]    
\includegraphics[width=6.5cm]{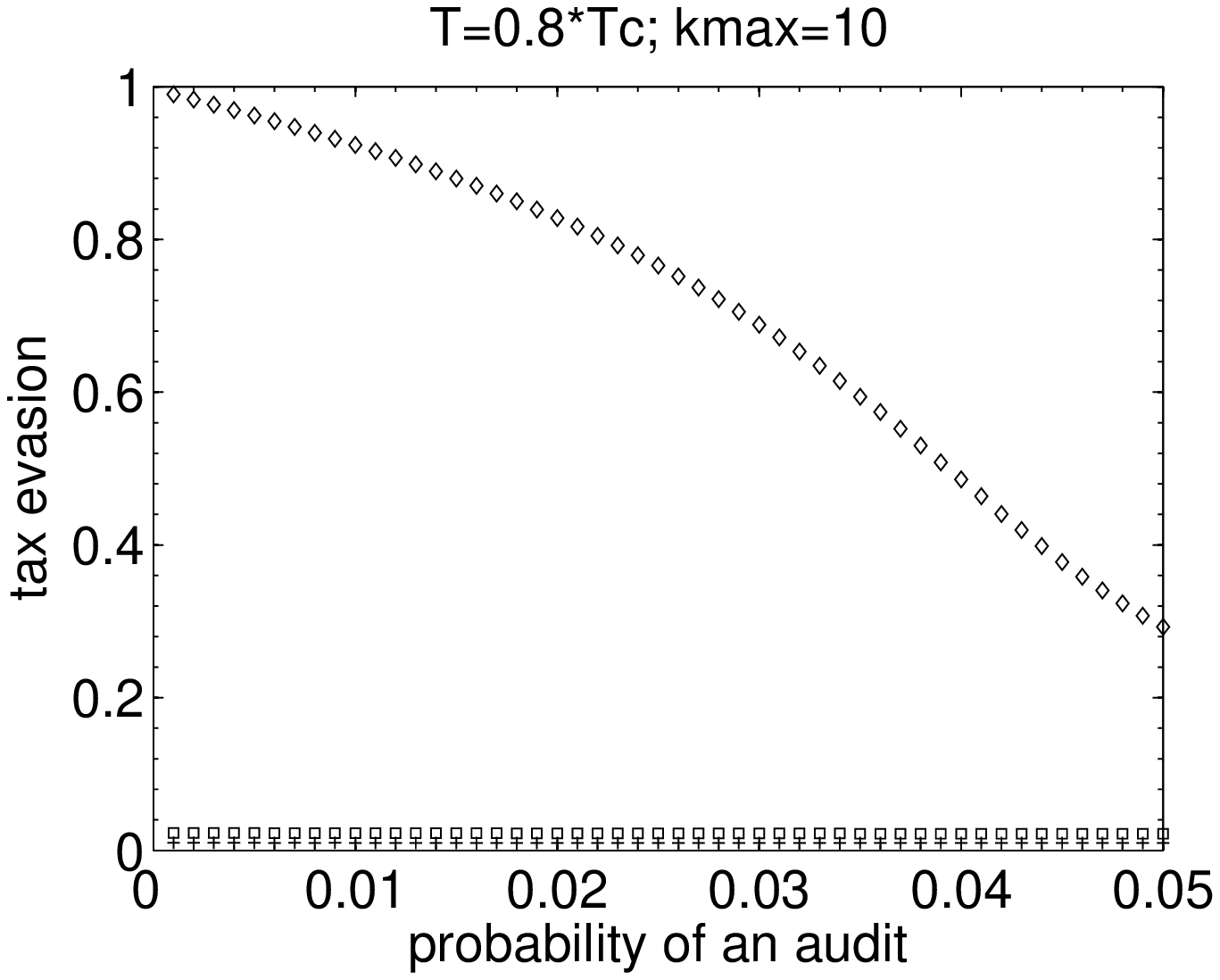}
\hspace{0.25cm}
\includegraphics[width=6.5cm]{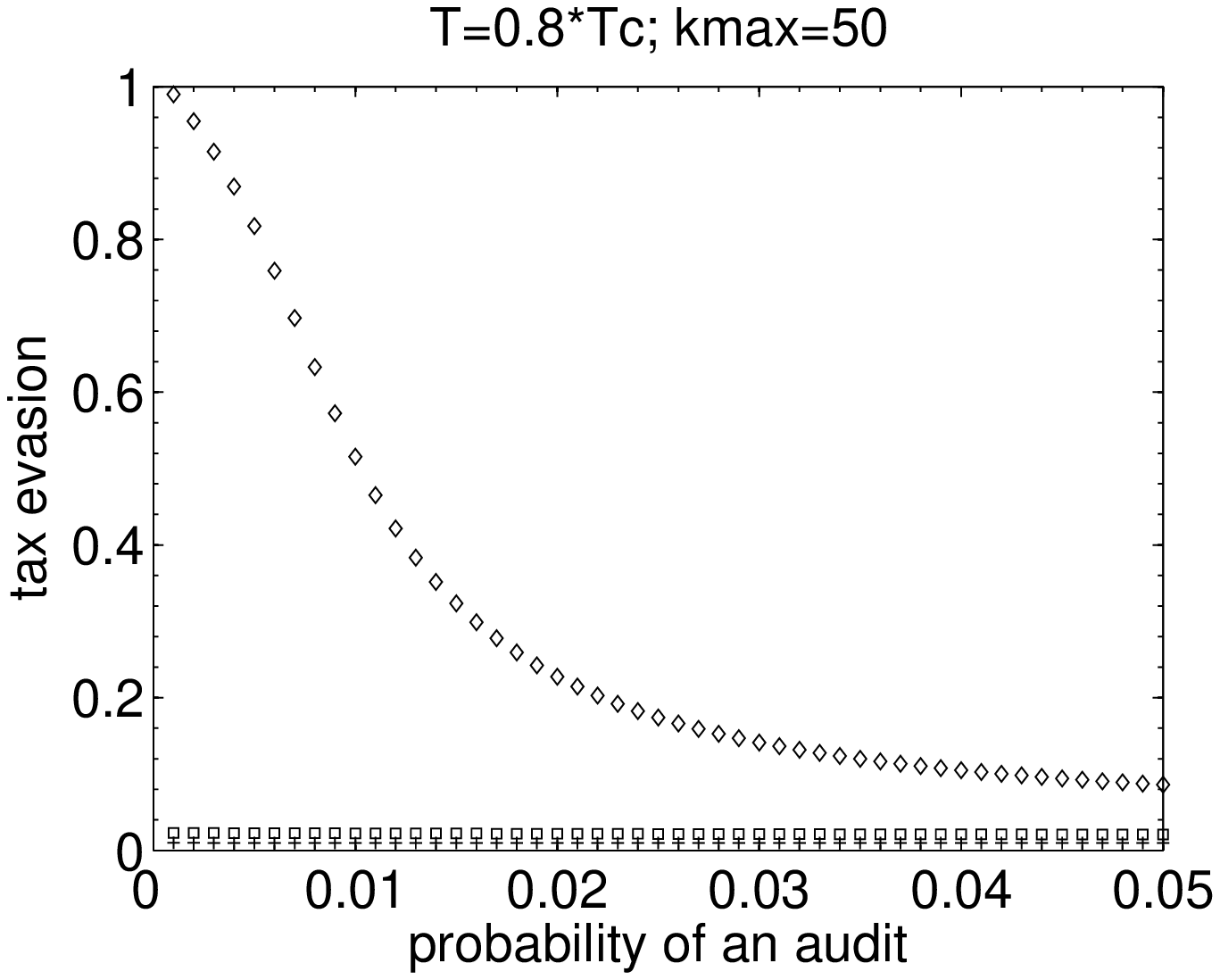}
%ende
\end{tabular}
\end{center}

\enlargethispage{\baselineskip}
\enlargethispage{\baselineskip}
\begin{minipage}[t]{\textwidth}
\vspace{0.25cm}
Figure 2: The panels on the left imply $k\in[0,10]$ and those on the right $k\in[0,50]$. They 
illustrate the equilibrium levels of tax evasion for different probabilities of an audit 
(the same applies to Figs. 3 and 4) in the square lattice network (cp. section 3).
\end{minipage}

%%%%%%%%%%%%%%%%%%%%%%%%%%Abbildung 3%%%%%%%%%%%%%%%%%%%%%%%%%%
\newpage
%\vspace*{-2.5cm}
\begin{center} 
\begin{tabular}{c}
\includegraphics[width=6.5cm]{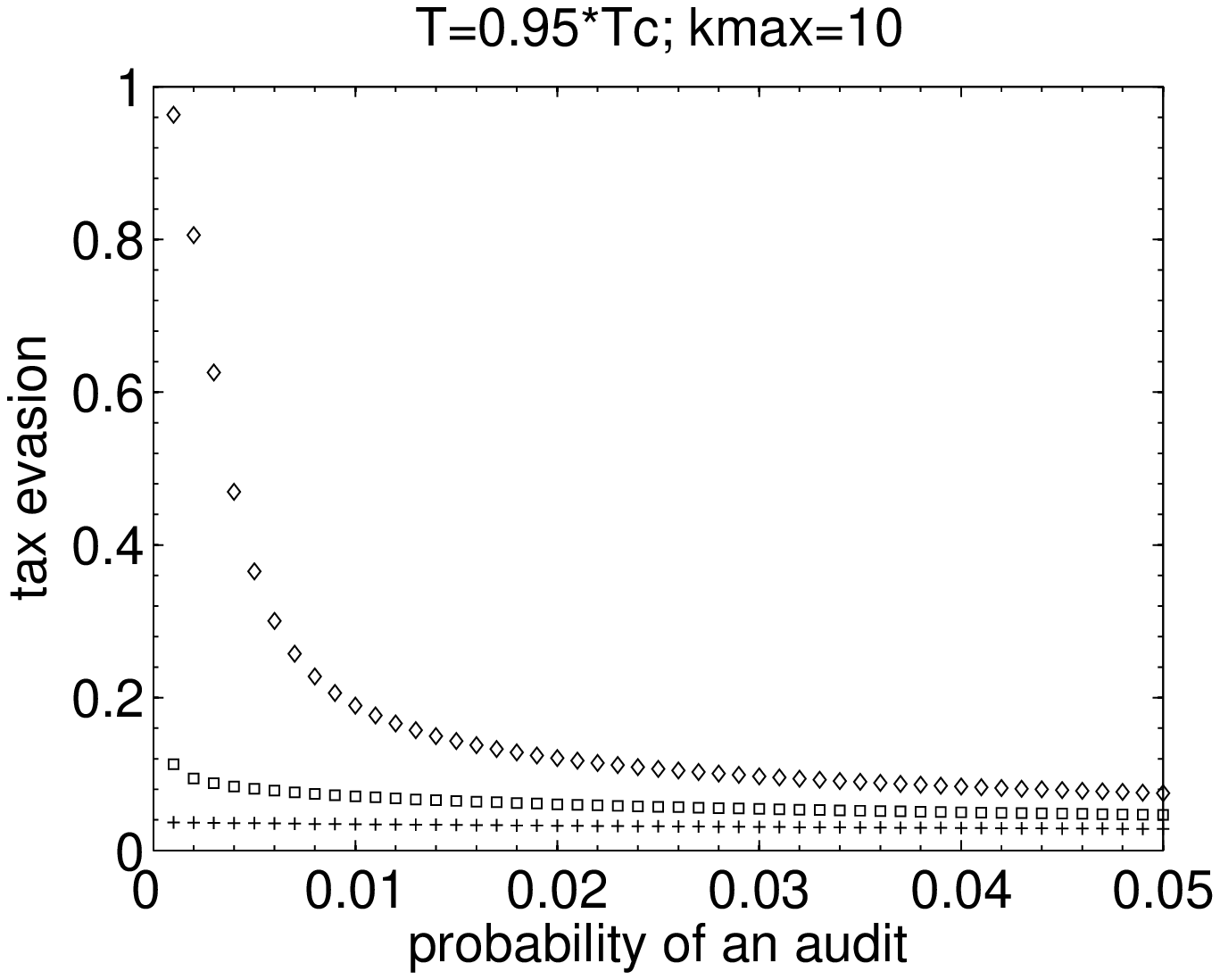}
\hspace{0.25cm}
\includegraphics[width=6.5cm]{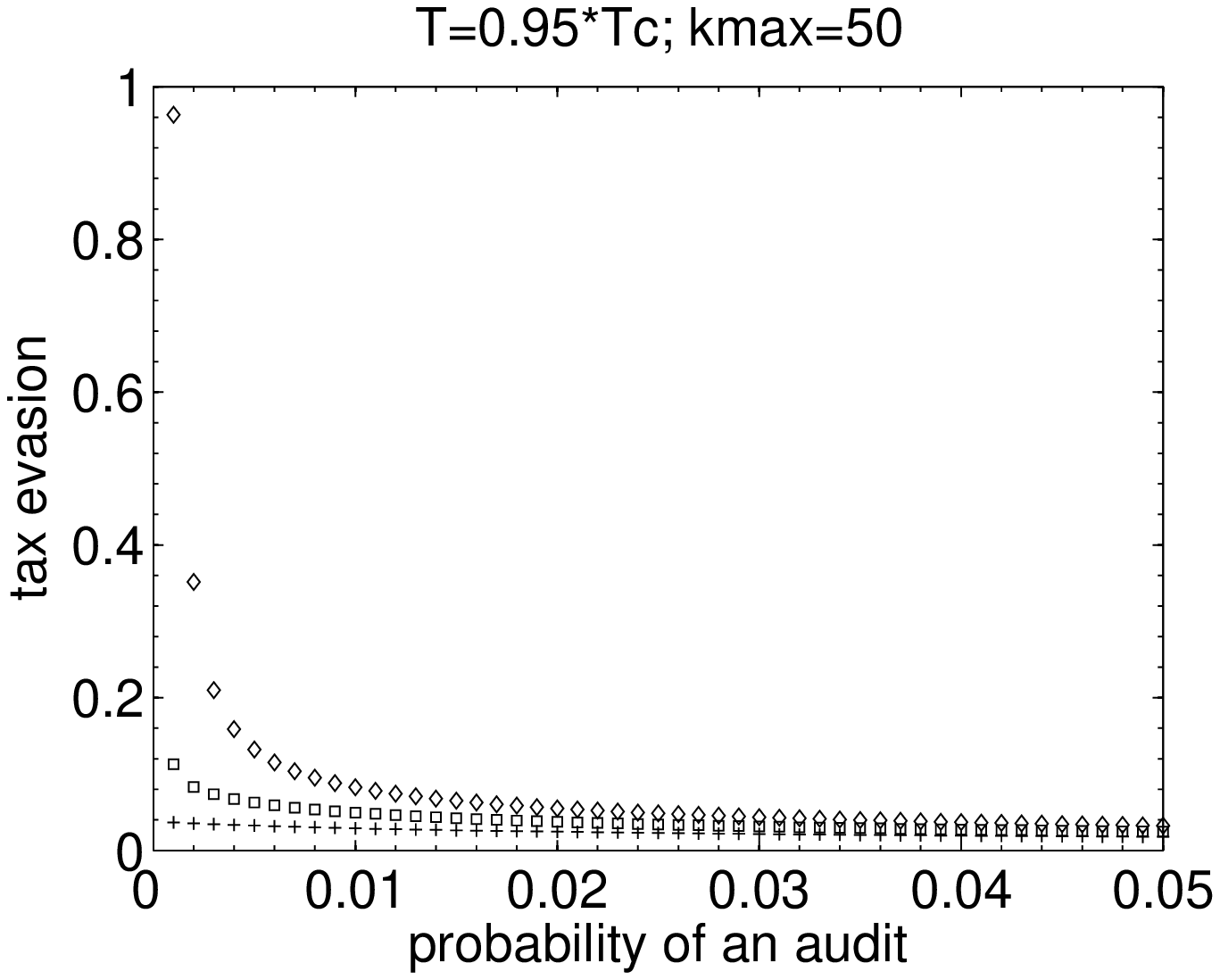}\\[0.25cm]  
\includegraphics[width=6.5cm]{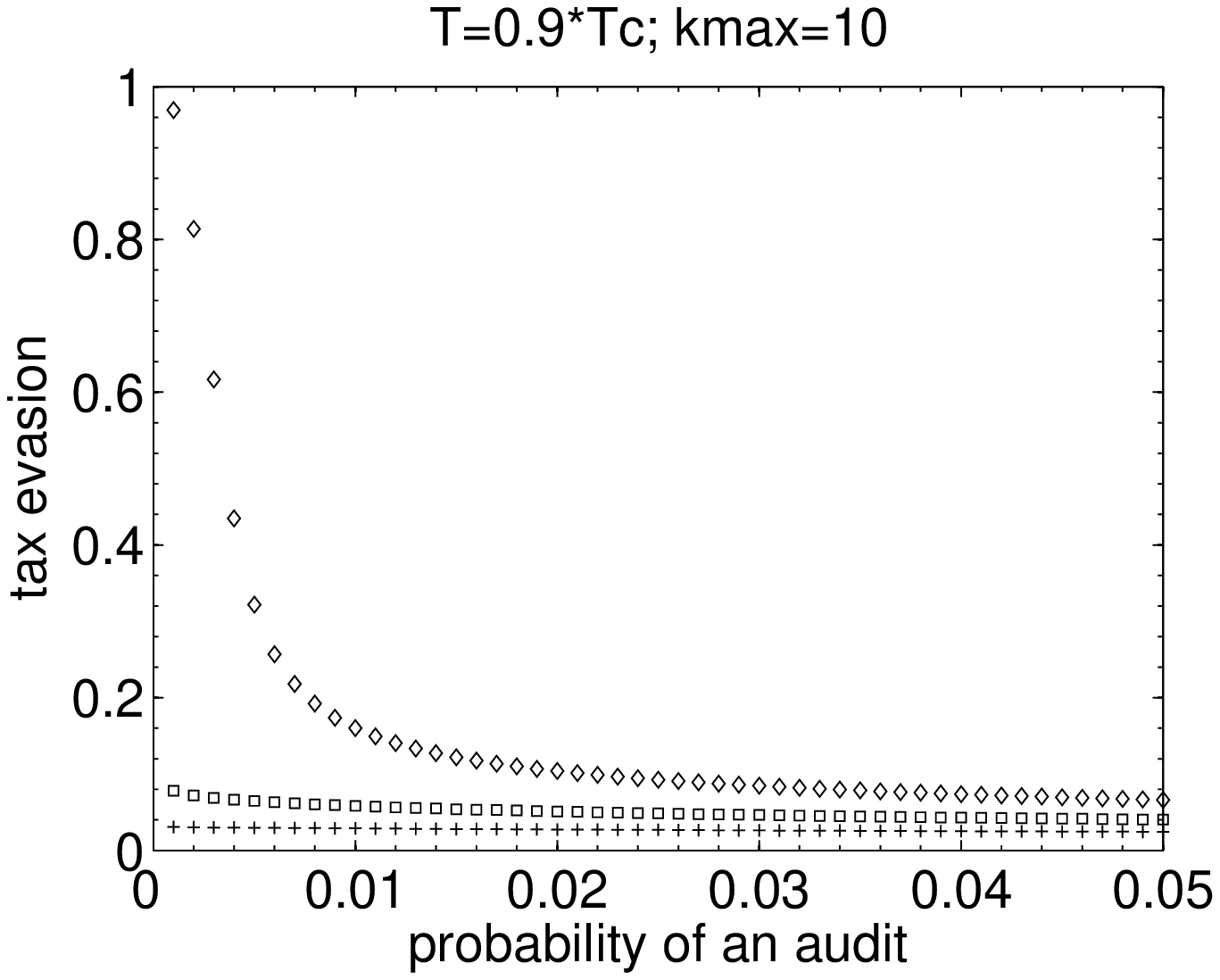}
\hspace{0.25cm}
\includegraphics[width=6.5cm]{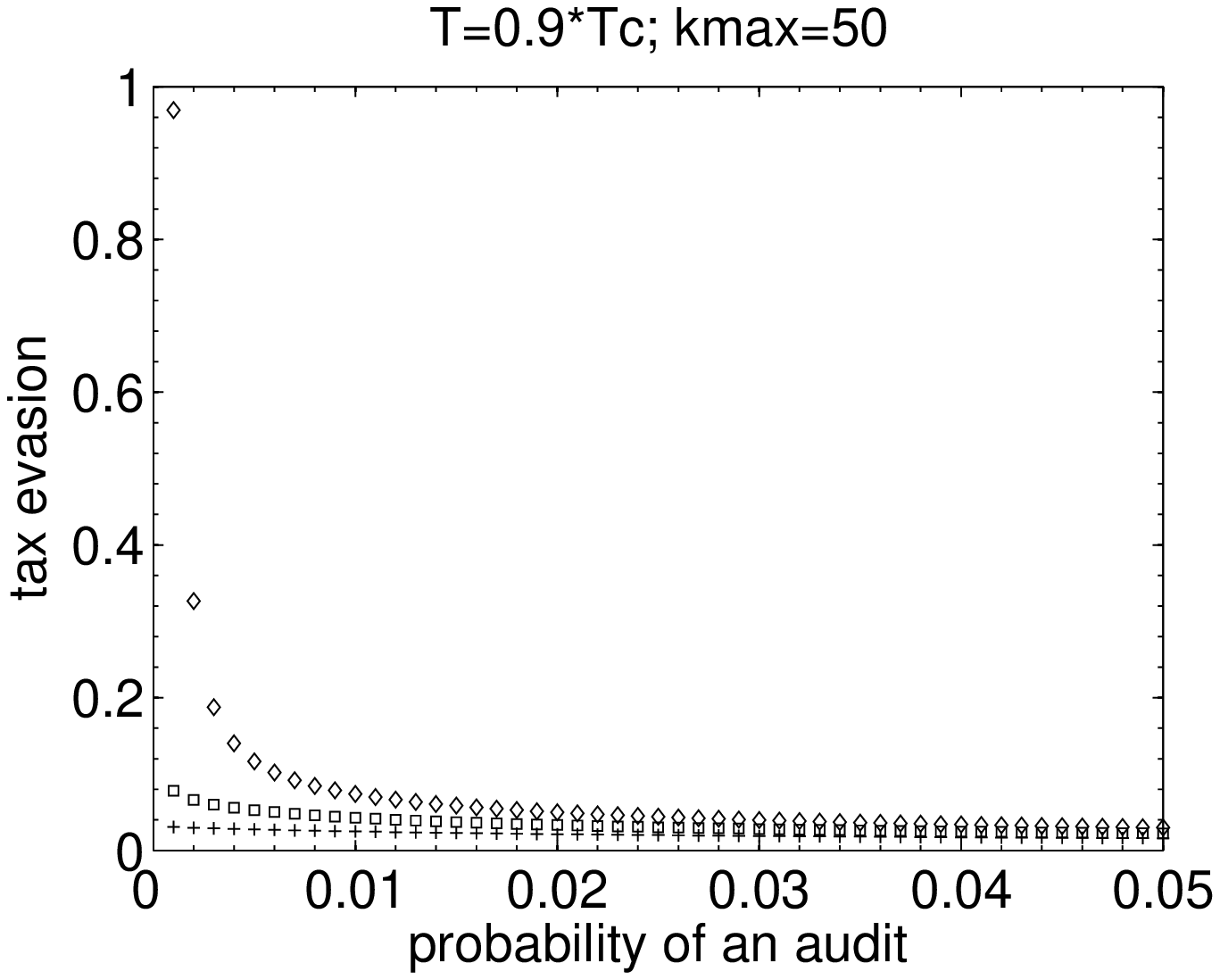}\\[0.25cm]    
\includegraphics[width=6.5cm]{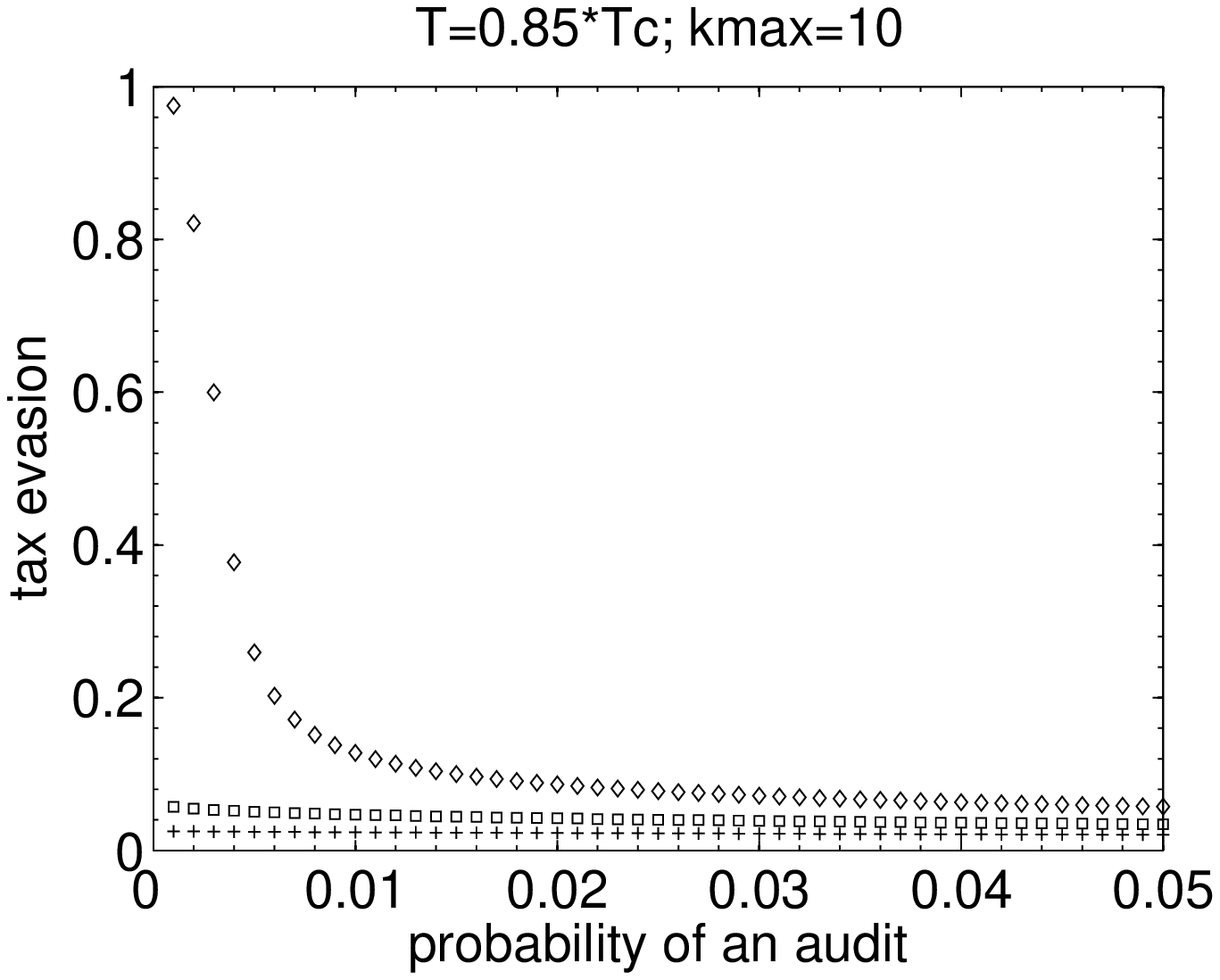}
\hspace{0.25cm}
\includegraphics[width=6.5cm]{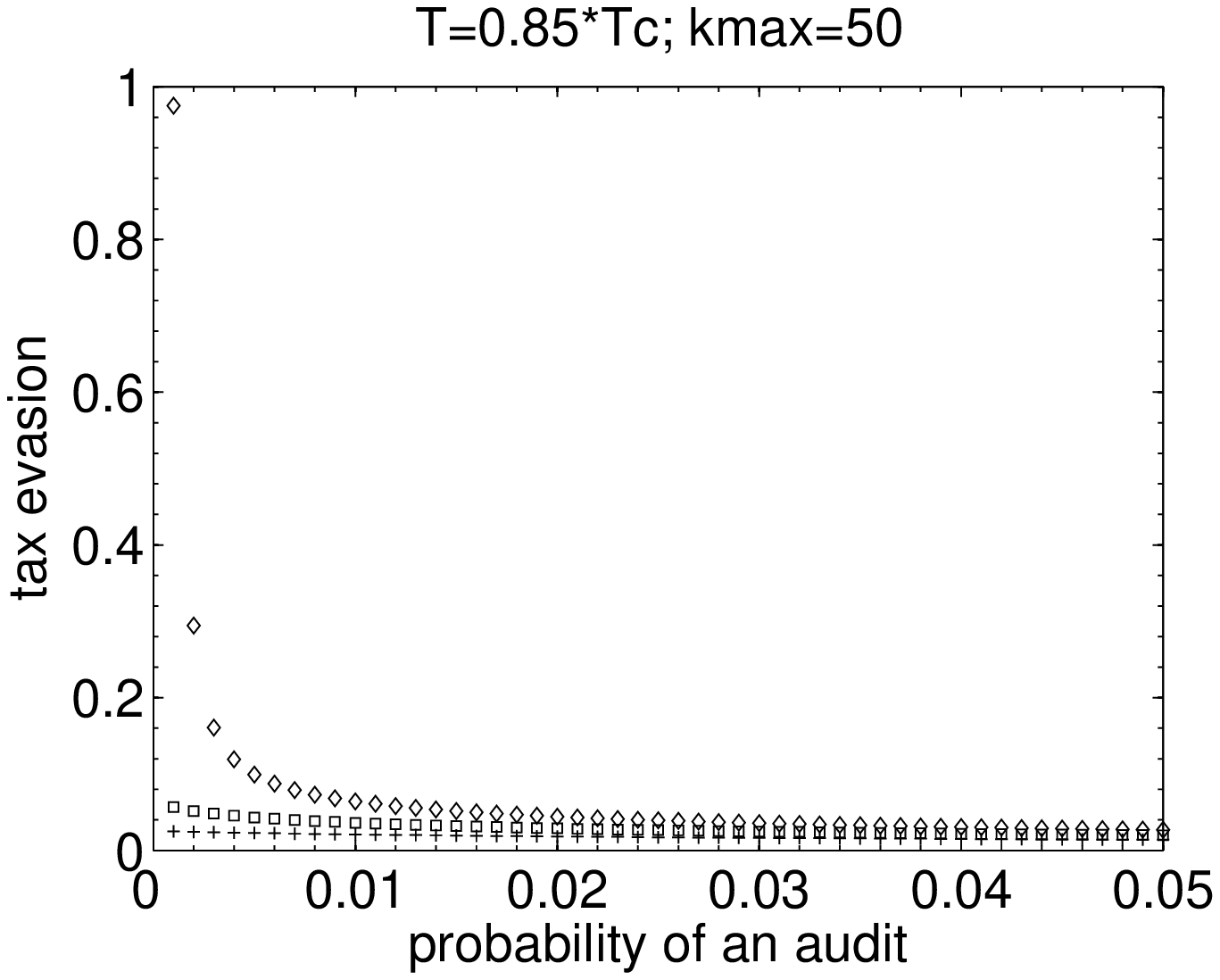}\\[0.25cm]    
\includegraphics[width=6.5cm]{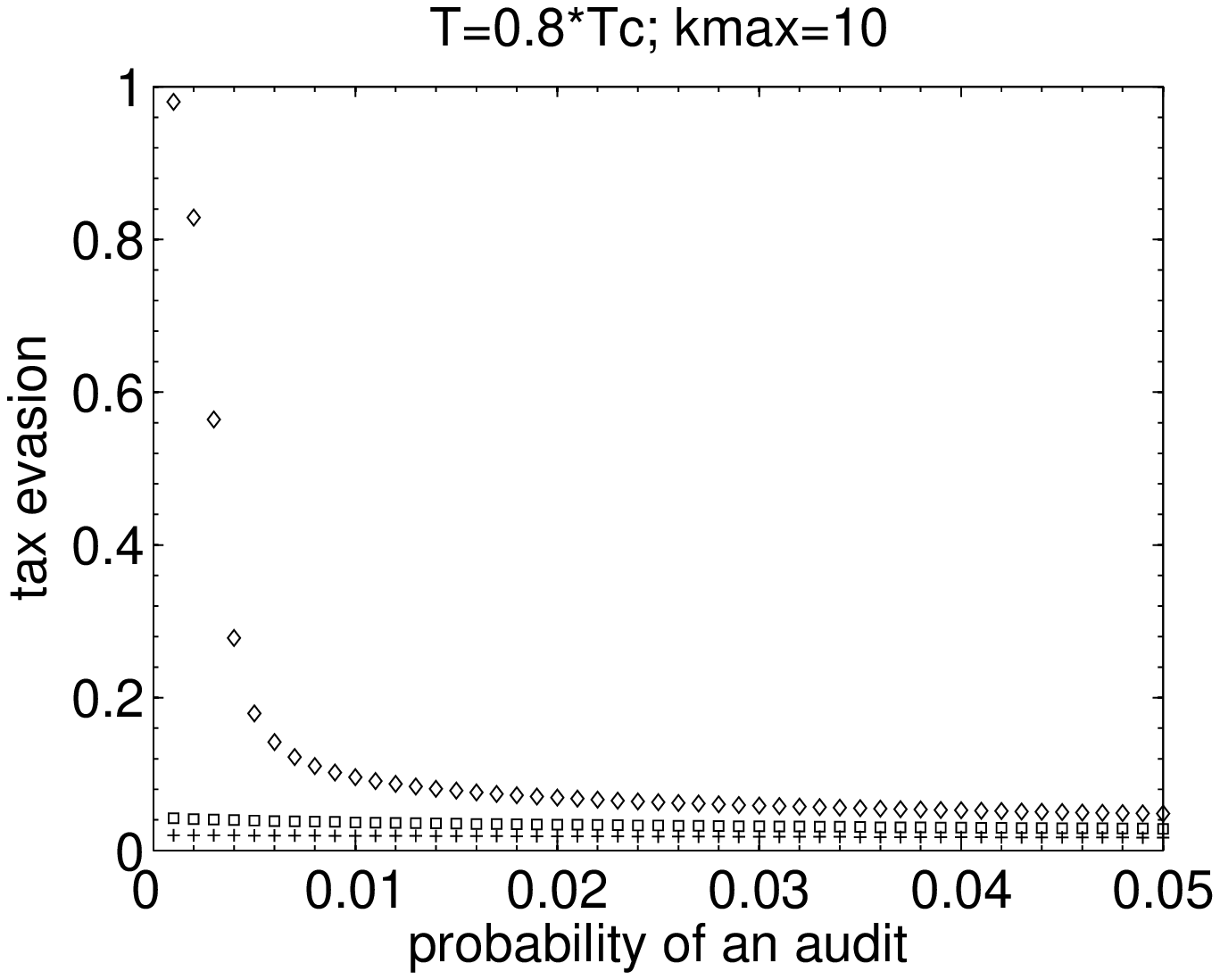}
\hspace{0.25cm}
\includegraphics[width=6.5cm]{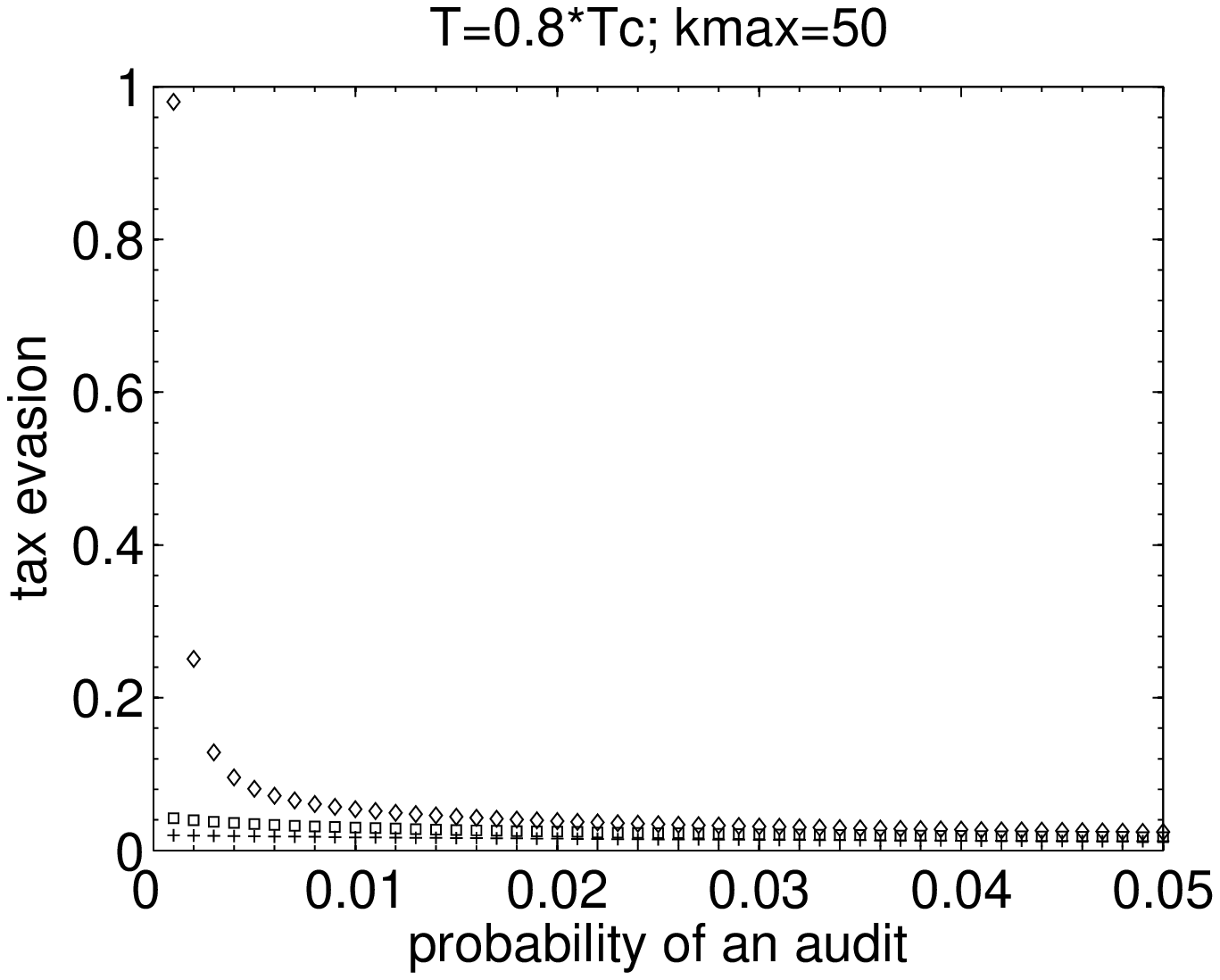}
%ende
\end{tabular}
\end{center}

\enlargethispage{\baselineskip}
\enlargethispage{\baselineskip}
\begin{minipage}[t]{\textwidth}
\vspace{0.25cm}
Figure 3:  The same simulation setting as in Figure 2 applies to the 
Voronoi-Delaunay lattice, which has a critical temperature of 
$T_c=3.802$. (A more detailed decription is given in section 4.)
\end{minipage}

%%%%%%%%%%%%%%%%%%%%%%%%%%Abbildung 4%%%%%%%%%%%%%%%%%%%%%%%%%%5
\newpage
%\vspace*{-2.5cm}
\begin{center} 
\begin{tabular}{c}
\includegraphics[width=6.5cm]{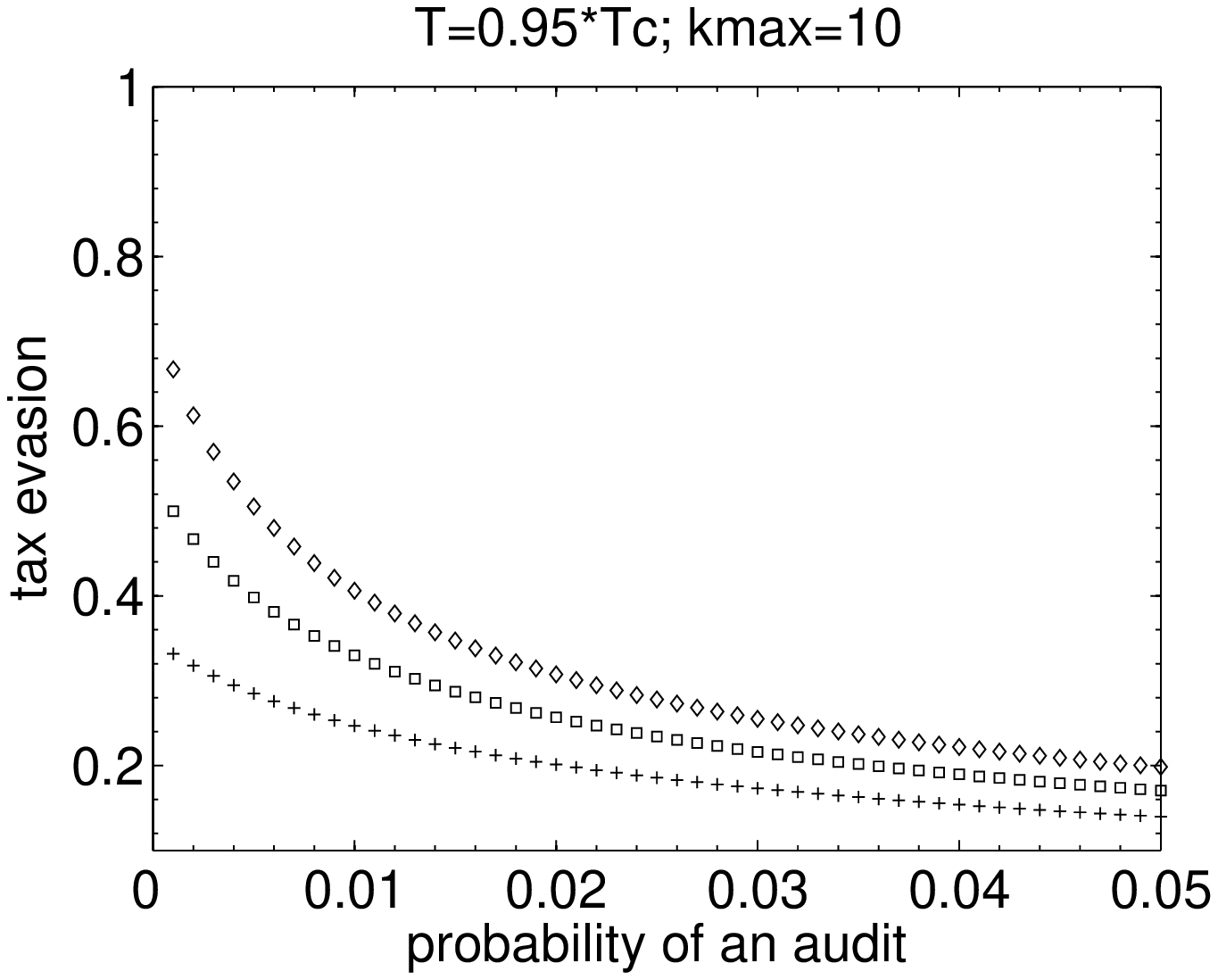}
\hspace{0.25cm}
\includegraphics[width=6.5cm]{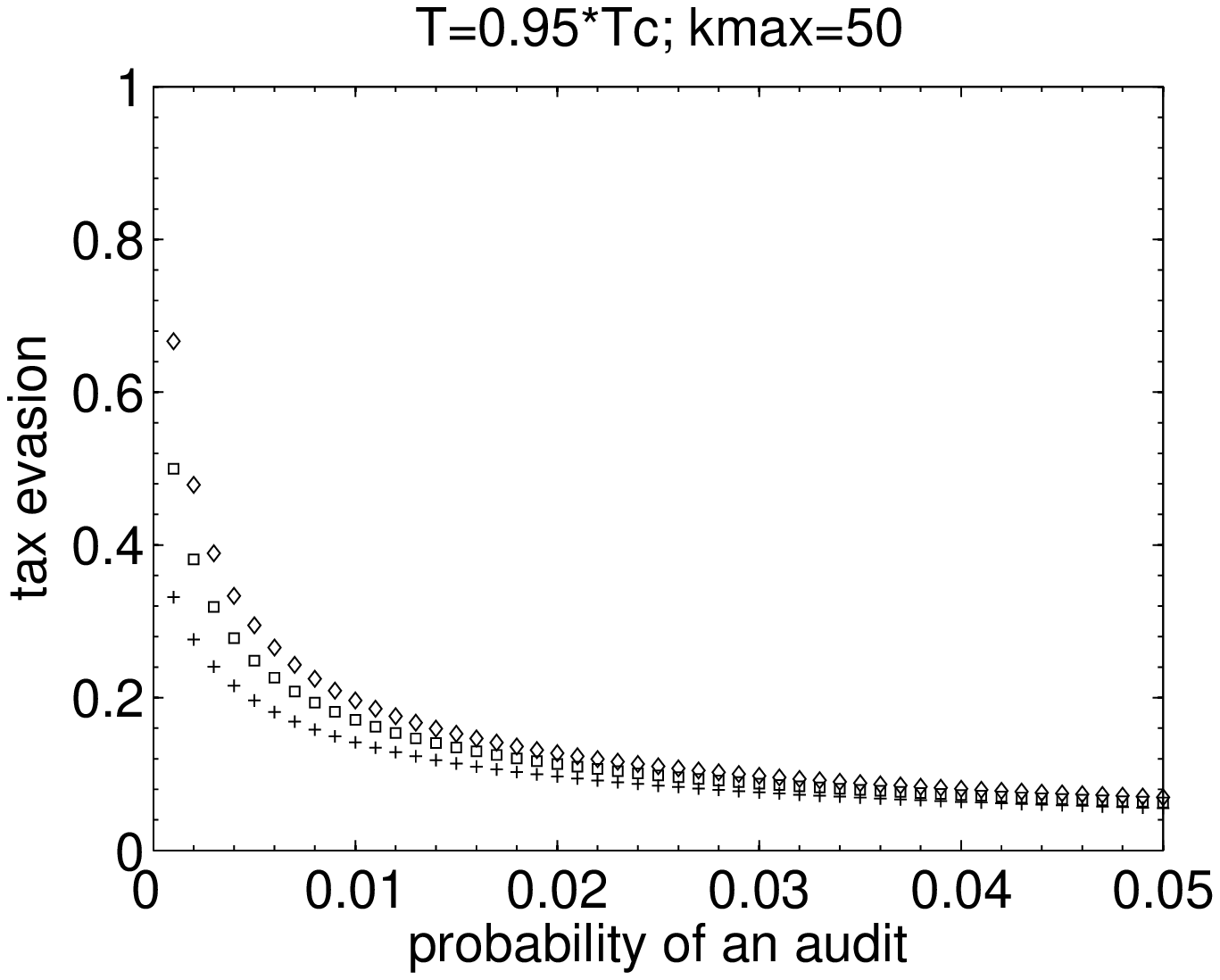}\\[0.25cm]  
\includegraphics[width=6.5cm]{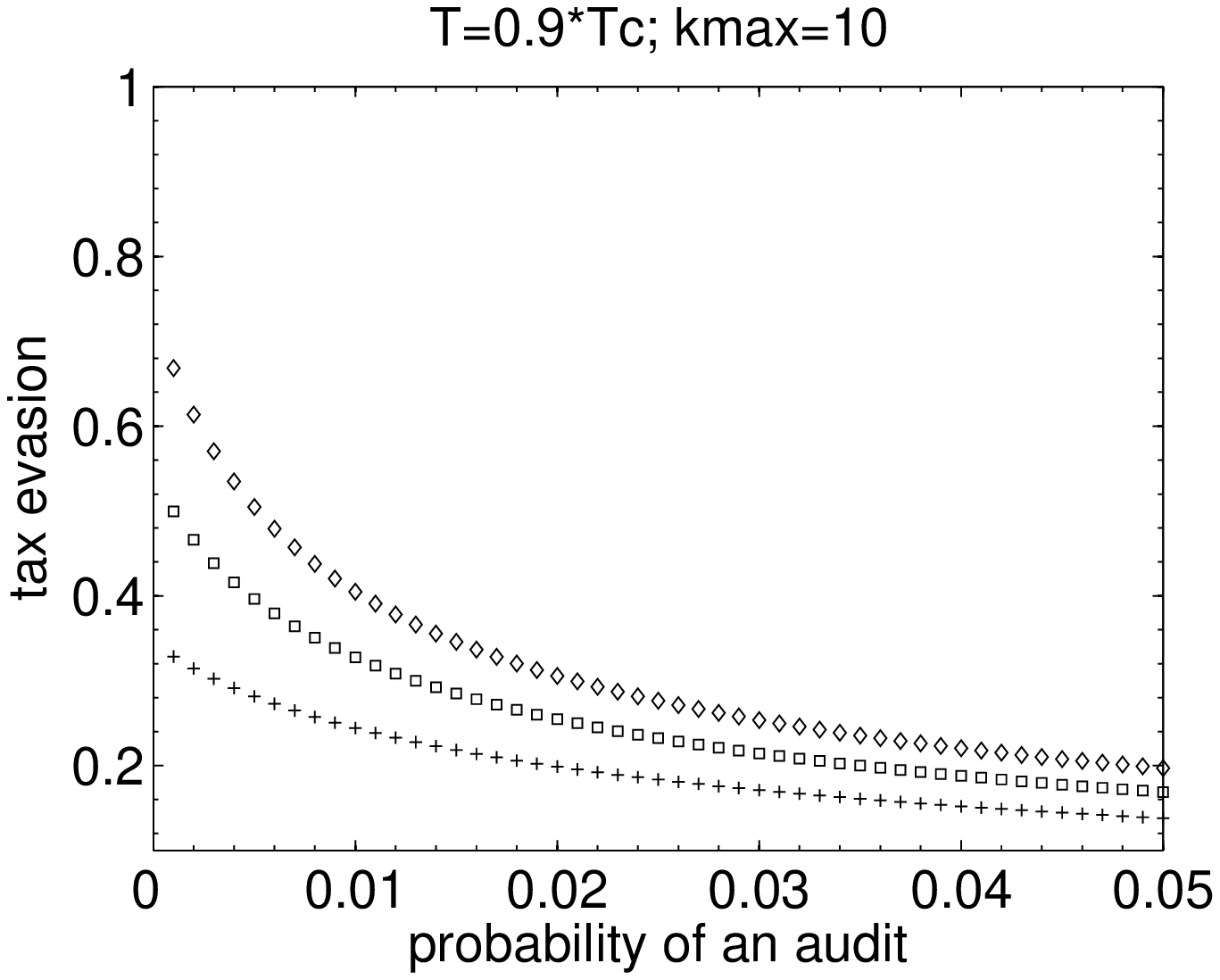}
\hspace{0.25cm}
\includegraphics[width=6.5cm]{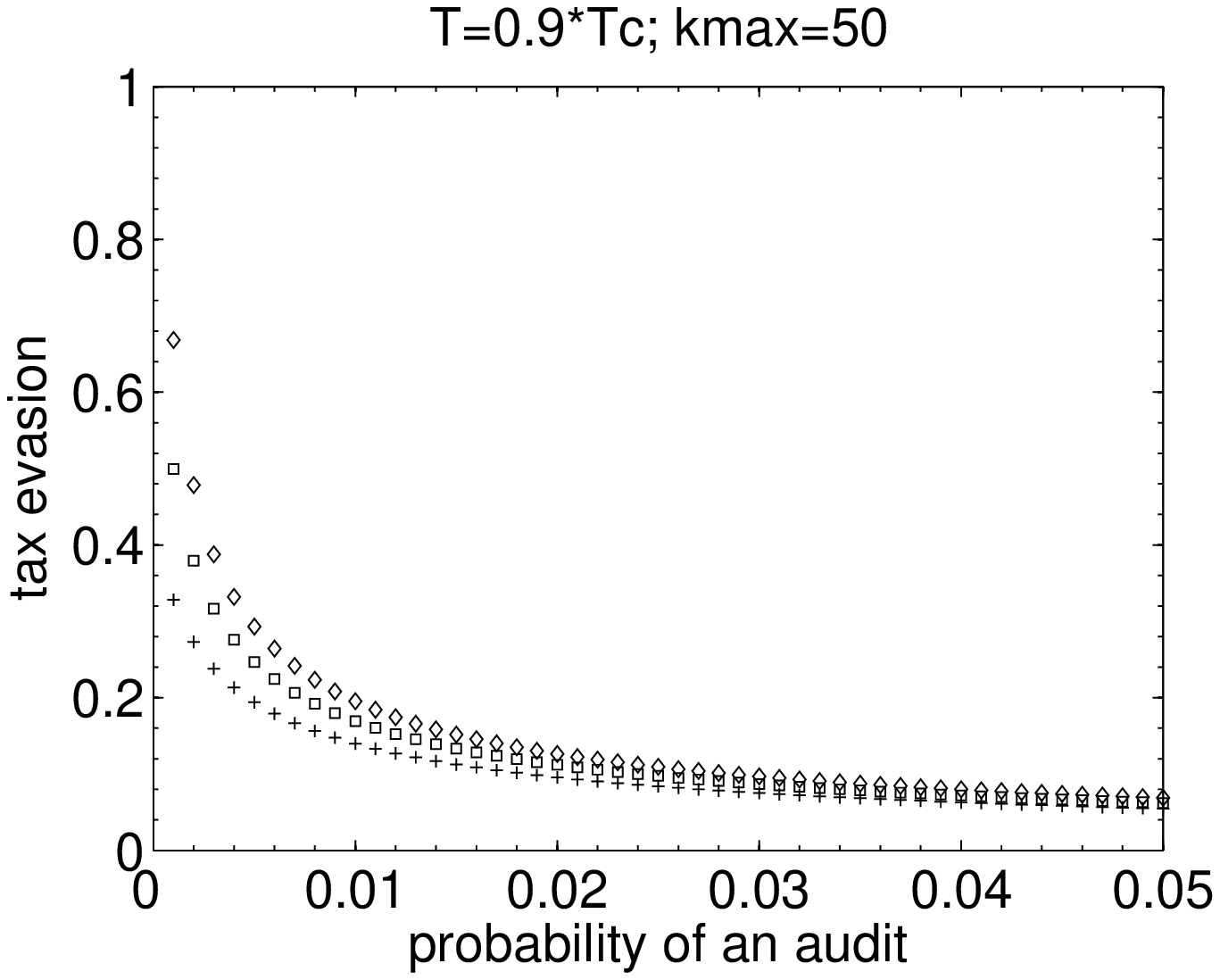}\\[0.25cm]    
\includegraphics[width=6.5cm]{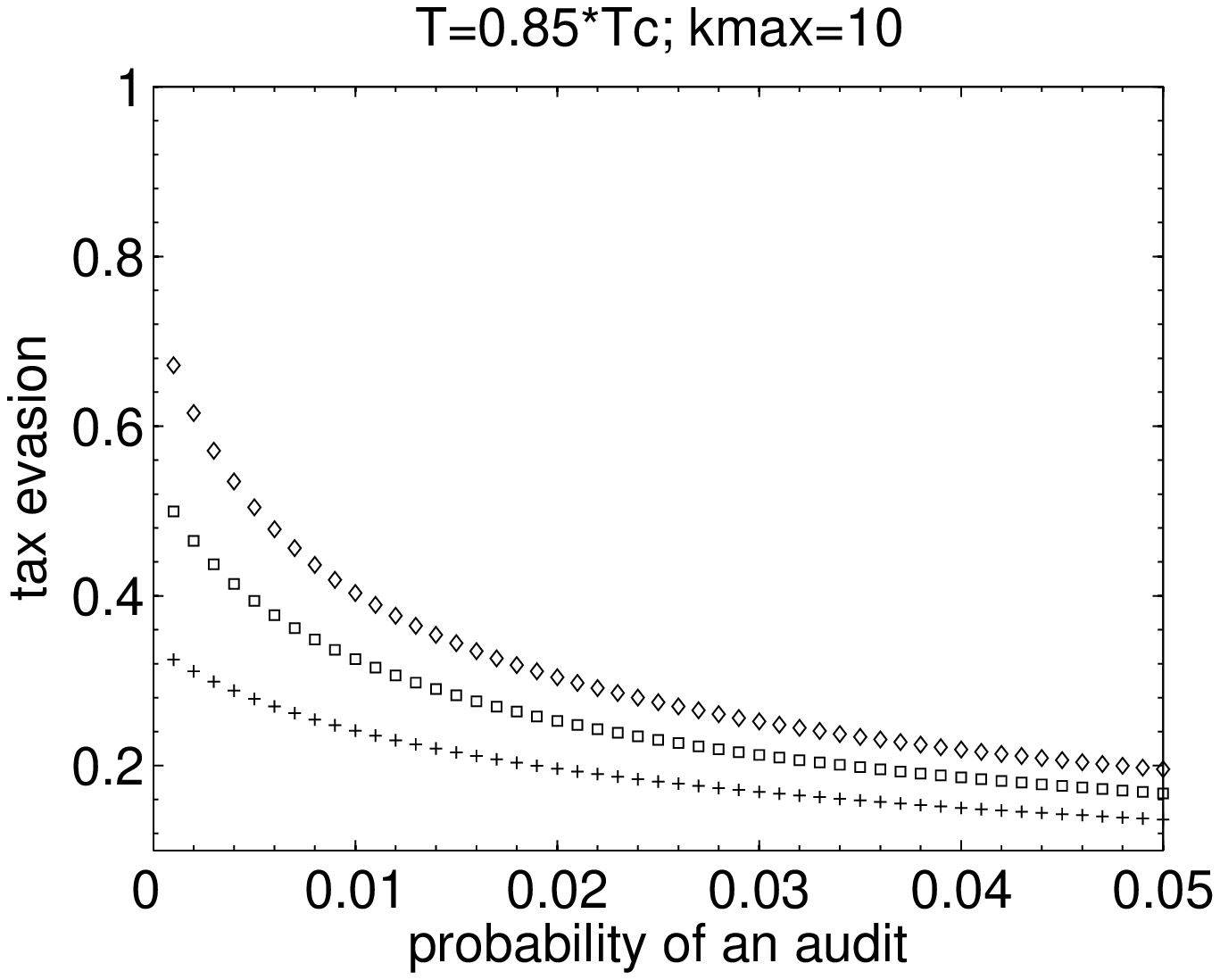}
\hspace{0.25cm}
\includegraphics[width=6.5cm]{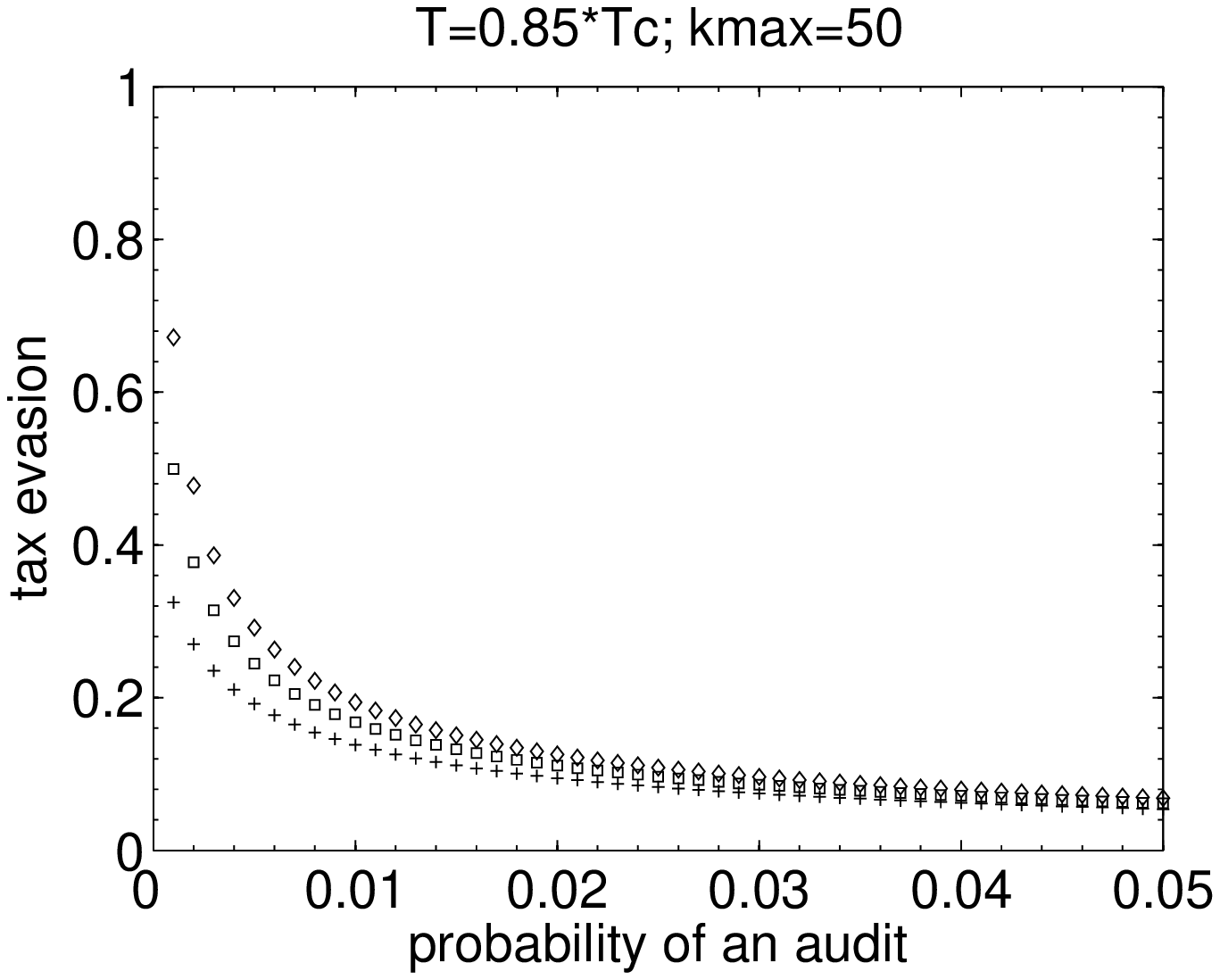}\\[0.25cm]    
\includegraphics[width=6.5cm]{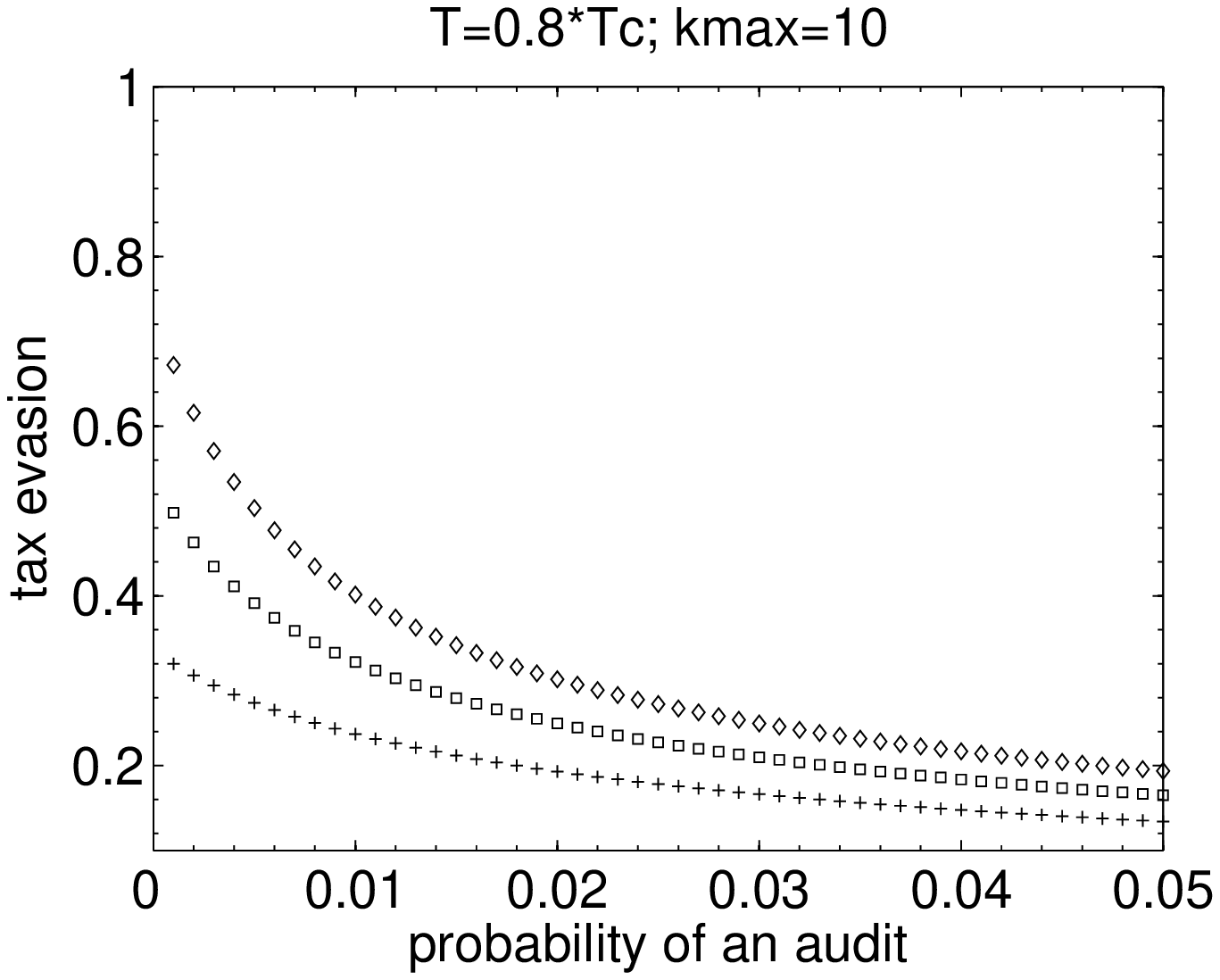}
\hspace{0.25cm}
\includegraphics[width=6.5cm]{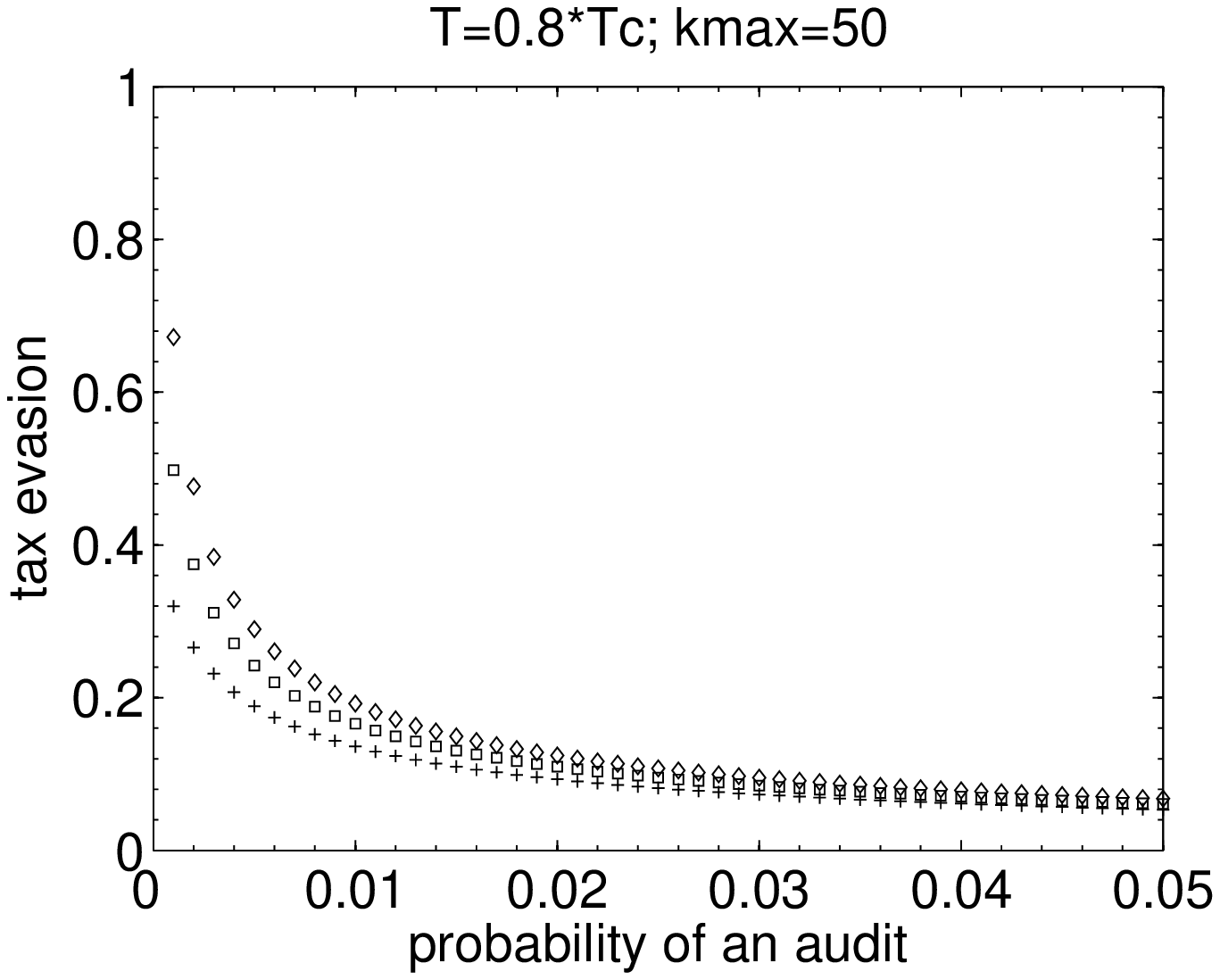}
%ende
\end{tabular}
\end{center}

\enlargethispage{\baselineskip}
\enlargethispage{\baselineskip}
\begin{minipage}[t]{\textwidth}
\vspace{0.25cm}
Figure 4:   The same simulation setting as in Figure 2 applies to the 
Barab\'asi-Albert network, which has a critical temperature of 
$T_c=m\cdot\mbox{log}(NSITES)/2$, where $m=4$ and $NSITES=1,000,000$. 
(A more detailed decription is contained in section 4.)
\end{minipage}

\end{document}